\documentclass{emulateapj}
\usepackage{apjfonts}
\usepackage{lscape}

\slugcomment{To appear in the Astronomical Journal}

\def\hst{{\sl HST}}
\def\iras{{\sl IRAS}}
\def\msun{M$_{\odot}$}
\def\ipol{$I_{\rm p}$}
\makeatletter
\def\ale{\mathrel{\mathpalette\gl@align<}}
\def\age{\mathrel{\mathpalette\gl@align>}}
\def\gl@align#1#2{\lower.6ex\vbox{\baselineskip\z@skip\lineskip\z@
\ialign{$\m@th#1\hfil##\hfil$\crcr#2\crcr\sim\crcr}}}
\makeatletter

\shorttitle{HST NICMOS Imaging Polarimetry of PPNs. II} 
\shortauthors{Ueta et al.}

\begin{document}
 
\title{HST/NICMOS Imaging Polarimetry of Proto-Planetary Nebulae II:\\ 
Macro-morphology of the Dust Shell Structure via Polarized Light}  

\author{%
Toshiya Ueta\altaffilmark{1},
Koji Murakawa\altaffilmark{2},
Margaret Meixner\altaffilmark{3}}

\altaffiltext{1}{Department of Physics and Astronomy, University of
Denver, Denver, CO 80208, USA; tueta@du.edu} 

\altaffiltext{2}{%
Max-Planck-Institut f\"{u}r Radioastronomie,
Auf dem H\"{u}gel 69, G-53121, Bonn, Germany}

\altaffiltext{3}{Space Telescope Science Institute, 
3700 San Martin Drive, Baltimore, MD 21218, USA}

\begin{abstract}
The structure of the dusty circumstellar envelopes (CSEs) of
 proto-planetary nebulae (PPNs) reveals the mass-loss history of
 these sources and how such histories may differ for elliptical (SOLE)
 and bipolar (DUPLEX) PPNs.
To study the PPN structures via dust-scattered linearly polarized
 starlight, we have compiled the imaging-polarimetric data for all 18
 evolved stars that have been obtained to date with NICMOS on-board the
 {\sl Hubble Space Telescope} (\hst).  
This alternative imaging technique provides a unique way to probe the
 distribution of dust grains that scatter light around evolved stars.
The new perspective gained from the imaging-polarimetric data has
 revealed several new aspects to the structures of PPNs.
Point-symmetry is a prevalent imaging-polarimetric characteristic
 resulting from the azimuthal density gradient in the CSEs.
Among these point-symmetric nebulae, three detailed morphological types
 can be differentiated by their polarized intensity, {\ipol}, and
 polarization strength, $P$. 
While the azimuthal density gradient is reversed above and below the
 equatorial plane in optically thicker bipolar nebulae, there is no
 gradient reversal in optically thinner elliptical nebulae.
The equatorial plane of the system defined by the integrated angle of
 polarization is not necessarily orthogonal to the axis of the apparent
 bipolar structure in the total intensity data.
\end{abstract}


\section{Introduction}

Proto-planetary nebulae (PPNs; e.g.\ \citealt{kwok93,vanwinckel03}) are
low to intermediate initial mass ($\sim 1 - 8$ \msun) objects
that are rapidly becoming planetary nebulae (PNs) after evolving off
the asymptotic giant branch (AGB).
A PPN consists of a post-AGB central star and a circumstellar envelope
(CSE) of gas and dust that is created by mass loss from the central star
during the AGB phase. 
When mass loss ceases at the end of the AGB phase, the CSE physically
detaches from the central star and keeps coasting away during the PPN
phase. 
Thus, PPNs preserve the AGB mass-loss history imprinted in the density
distribution of the CSEs.
Therefore, PPNs are the best astrophysical laboratories in which we 
investigate the nature of dusty mass loss during the AGB phase.

During the last decade, high-resolution imaging observations have
revealed a plethora of morphologies among PNs and PPNs (see review by
\citealt{balick02}).   
One of the most important tasks in the AGB stellar research is
therefore to determine the main physical factors that cause such a
great morphological diversity emerging from AGB mass-loss geometry,
which is highly spherically symmetric at the beginning of the AGB phase
(e.g.\ \citealt{habing93}).   
The observed morphological diversity suggests a number of shell shaping  
processes, and various mechanisms have been proposed to explain the
observed morphologies (\citealt{balick02}, and references therein). 
In all possible cases, however, the CSE structure must deviate from the
initial spherical symmetry.
Therefore, there may exist a universal mechanism through which the
required symmetry breaking is achieved, and thus, we have been
performing surveys to identify the mechanisms that generate the initial 
non-sphericity in the CSEs of AGB stars. 

Using mid-IR ($8-21\micron$) data, \citet{meixner99} have shown that 
all the extended PPNs in their sample are axisymmetric and fall into one
of the two morphological classes: toroidal (showing limb-brightened
peaks that suggest the presence of a dusty torus) or core/elliptical
(showing an unresolved emission core surrounded by a low surface
brightness elliptical nebula).   
They concluded that the last stage of AGB mass loss is short-lived and
inherently toroidal (i.e., equatorially-enhanced), probably coinciding
with a period of enhanced mass loss known as the ``superwind'' phase
\citep{renzini81,iben83}. 

In an optical imaging survey of PPN candidates using the Hubble Space
Telescope ({\hst}), \citet{ueta00} have detected elongated reflection
nebulosities whose morphologies appear to bifurcate depending on the
degree of the equatorial density enhancement in the CSE.
SOLE (Star-Obvious Low-level Elongated) PPNs show a bright central
star embedded in a faint, elliptically-elongated nebulosity, while DUPLEX
(DUst-Prominent Longitudinally EXtended) PPNs exhibit a remarkable
bipolar structure with a dust lane completely or partially obscuring the
central star.  
These optical morphological classes appear to possess a one-to-one
correspondence to the mid-IR counterparts (core/elliptical $\approx$
DUPLEX and toroidal $\approx$ SOLE).   

Subsequent radiative transfer modeling of some selected sources (e.g.\
\citealt{meixner97,ueta01,meixner02,meixner04}) have shown that (1) the 
observed PPN axisymmetry emerges during a brief stage of an enhanced
mass loss ($\sim 3 \times 10^{-5}$ {\msun} yr$^{-1}$) towards the end of
the AGB phase, (2) the enhanced mass-loss rate into the equatorial
plane is very much higher ($\age 10^{3}$) than that into the polar
directions, and (3) the differences between SOLE and DUPLEX PPNs stem
from the physical distinction between the two types and cannot be  
attributed solely to the inclination angle effects. 
Furthermore, it appears that SOLE and DUPLEX PPNs develop respectively
from low and high mass progenitors.
This agrees with a recent discovery of mostly bipolar structures among
OH/IR-selected sources, which are located close to the galactic plane
(i.e., more massive), observed by {\hst} \citep{sahai04}.  

Imaging polarimetry is an alternative technique to probe the CSE
structure, in which the dust-scattered linearly polarized component of
light can be separated from the unpolarized component of light.
Gledhill and collaborators \citep{gledhill01,gledhill05} have found,
through UKIRT imaging-polarimetric surveys of PPN candidates, that the
axisymmetric structure of the PPN shells fall into one of the two
classes, {\sl shell} and {\sl bipolar}, depending on the amount of dust
in the CSE.  
These classes respectively correspond to SOLE and DUPLEX PPNs.
The results of the previous mid-IR and optical surveys have thus been 
confirmed.

We have performed near-IR imaging polarimetry with {\hst} to investigate
the dust distribution in SOLE PPNs by making use of the polarized
intensity ({\ipol}\/) maps.     
Imaging polarimetry is especially powerful for this purpose, because  
the technique isolates dust-scattered light arising from the CSE and
permits us to probe the CSE structure even in the vicinity of the bright
central star; CSE structure is often obscured in total intensity ($I$)
maps, which are affected by the enormous point-spread-function (PSF) of
the central star.    
Since {\ipol} depends only on the Stokes $Q$ and $U$, which are free
from the unpolarized component by definition, it can be extracted  
even from the PSF-affected data.
We have successfully demonstrated that (1) SOLE PPNs consist of a hollow 
spheroidal shell which has a built-in equatorial density enhancement and
(2) the variable degree of the equatorial density enhancement (i.e., the
optical depth of the CSE) would result in a variety of apparent PPN 
structure even among this specific group of optically thin PPNs
(\citealt{ueta05}, hereafter Paper I).   

In the present work, we extend our investigation of the CSE structure
via {\ipol} into the realm of DUPLEX PPNs. 
We apply the technique of imaging polarimetry in investigating the
structure of the {\sl bipolar lobes}, in which the optical depth is
generally low enough for the method to work. 
To date, there have been observations of 18 evolved stars by
{\hst}/NICMOS in the polarimetric mode.
Only a small portion of the entire data set has been published, and the
{\ipol} data have rarely been utilized to study the CSE dust distribution.
Therefore, we analyze all the existing {\hst}/NICMOS imaging
polarimetric data, especially in {\ipol}, obtained from the archive in
order to obtain more insights into the general mechanism(s) responsible
for the structure development in the AGB wind shells.
Below, we describe the data set and reduction procedure briefly (\S 2),
present and discuss the results both individually and in groups (\S 3)
and give a summary (\S 4). 

\section{Observations and Data Reduction}

We have collected NICMOS imaging-polarimetric data for 13 evolved stars
from the {\hst} archive and combined them with our own data for 4 SOLE
PPNs that have already been published (Paper I).
When PSF correction is necessary, we have used data for two PSF
calibration sources, HD 12088 and BD$+$32$^{\circ}$3739, respectively in
the NIC1 and NIC2 band. 
Table \ref{data} summarizes the major observing parameters for all
observations.  
Unfortunately, the data for IRAS 02168 and IRAS 10131 (in the POL-L
band) are affected by severe saturation and not included in the
subsequent analysis.
In the following, we use the IRAS designation of the targets to refer to
each object except for AFGL 2688, which does not have an IRAS
designation. 

The standard set of NICMOS pipeline routines, CALNICA (Version 3.1), in
IRAF/STSDAS\footnote{%
IRAF is distributed by the NOAO, which is operated by AURA under
cooperative agreement with NSF. 
STSDAS is a product of STScI, which is operated by AURA for NASA}
has been used in data reduction.  
Additional care has been taken to remove the pedestal effect (a variable
quadrant bias) and the Mr.\ Staypuft (the amplifier ringing and
streaking due to bright targets; \citealt{nicmoshandbook}).  
When multiple frames are combined, we have used the drizzle package
\citep{dithering} in STSDAS.
The resulting images maintain their original pixel scales of
$0\farcs043$ pixel$^{-1}$ and $0\farcs075$ pixel$^{-1}$ respectively in
the NIC1 and NIC2 band.

Derivation of the Stokes parameters is done via the matrix inversion
method \citep{hines00,nicmoshandbook}.
However, the coefficients are different for data taken before and after
the servicing mission 3B (SM3B) in 2002 March, during which the NICMOS
cooling system was installed. 
This resulted in two sets of the photometric conversion factors 
for the pre- and post-SM3B data.
We have used these coefficients (listed in Table 2 of Paper I) where
appropriate.  
A more detailed account of data reduction is found in Paper I.

\section{Results and Discussion}

\subsection{Results}

The imaging-polarimetric maps of evolved stars are presented in a
uniform format, following the scheme in Paper I (Figs.\ \ref{04395} to
\ref{2688l}: Readers are encouraged to look at the color images
available on the electronic version of the paper.). 
In each figure, we show (a) the total intensity, $I$, map, (b) the
polarized intensity, {\ipol}, map, (c) the polarization strength, $P$,
map and (d) the polarization vector position angle (P.A.), $\theta$,
map, respectively from top to bottom.  
When the PSF-subtracted data are available, they are also displayed
in the right column (panels e--h) in addition to the original data in the
left panels, except for IRAS 10131 and AFGL 2688 for which
the two-epoch data are presented (the earlier/later epoch on the
left/right). 

The results are also summarized in Table \ref{results}, which lists the
measured coordinates and $I$ and {\ipol} fluxes of the object, the
maximum $P$ ($P_{\rm max}$), the mean $P$ ($\left<P\right>$) and its
standard deviation, and description of the {\ipol} structure.   
The listed coordinates are the center of the polarization vector pattern 
that is determined by minimizing the sum of the square of the
distance from the presumed center to a line orthogonal to the
polarization vectors.
The pattern center is initially assumed to be the $I$ peak (when seen),
and the center finding process is iterated until the shift between the
previous and current centers becomes smaller than the numerical accuracy
of the analysis.    
The fluxes are determined by integrating the surface brightness over an 
aperture in which there is more than one $\sigma_{\rm sky}$ detection.
$P_{\rm max}$ and $\left<P\right>$ plus its standard deviation are
determined by using an aperture which is delineated by the
well-established nebula edge or an aperture in which there is more than
three $\sigma_{\rm sky}$ detection. 
The {\ipol} structure is described by the overall shape, dimensions
(typically major and minor axis lengths), and P.A.\ measured from  
the north to the east.

To compare the present results with the previous unresolved polarization
measurements, we have also performed aperture polarimetry and summarized
the results with the measurements of Paper I in Table \ref{aperture}.
We have done aperture photometry on the three polarizer images and then
used the measured fluxes from each polarizer image to compute the
integrated $P$ and $\theta$.
We have varied the aperture size from $\times 1$ to $\times 8$ of FWHM  
in $I$ to determine how the resulting values of $P$ and $\theta$ behave
as a function of the aperture size.  
In general, $P$ and $\theta$ become stable and converge to a single
value when the aperture size is equal to or larger than five times the
FWHM size in $I$.  
Thus, the aperture size is defined to be the minimum aperture size 
(an integer multiple of FWHM in $I$) over which the integrated $P$ and  
$\theta$ are generally stable.  
The quoted errors in $P$ (0.2 to $4.0\%$) and $\theta$ (at most
$6^{\circ}$) are thus the maximum deviation in our measurements when the 
aperture size is varied beyond the adopted aperture size.
These values are consistent with the results of more thorough analyses
done by \citet{hines00}\footnote{In verifying the matrix coefficients by
measuring polarization of both polarized and unpolarized standards,
their measurements were consistent within $0.3\%$ in $P$ and $5^{\circ}$
in $\theta$} and 
\citet{batch06}\footnote{Possible instrumental polarization would make
the matrix coefficients uncertain and was found to cause difference of
$0.1\%$ in $P$ and $3^{\circ}$ in $\theta$.}. 

We have adopted the integrated $\theta$ as a measure of the orientation
of the equatorial plane because of the following reason. 
When the optical depth in the equatorial region is very high ($\tau
\gg 1$), there is little {\ipol} because of the dilution effects due to 
multiple scattering. 
In this case, the integrated $\theta$ tends to be parallel to the 
equatorial plane because of the orthogonal alignment of the scattering
plane with respect to the equatorial plane (e.g.,
\citealt{wh93,gledhill05}).  
However, when the optical depth is marginally high ($\sim 1$),
$\theta$ in the equatorial region can be orthogonal to the equatorial 
plane due to scattering at the walls of the bipolar lobes (e.g.,
\citealt{jj91,tdg94,gledhill05,oppen05}).    
Even in this case, the integrated $\theta$ tends to be parallel to the
equatorial plane, because vectors in the lobes away from the equatorial
region would still dominate. 
Therefore, the integrated $\theta$ is more appropriate as a
measure of the orientation of the equatorial plane, and is expressed as 
$\theta_{\rm eq}$ in the following.
Note, however, that $\theta_{\rm eq}$ is defined to be the angle
perpendicular to the orientation of the equatorial plane so that the
comparison between $\theta_{\rm eq}$ and P.A.\ of the {\ipol} surface
brightness distribution can be done intuitively. 

\subsection{Individual Sources}

\subsubsection{IRAS 04395$+$3601 (AFGL 618)}

\begin{figure}
\epsscale{.88}
\plotone{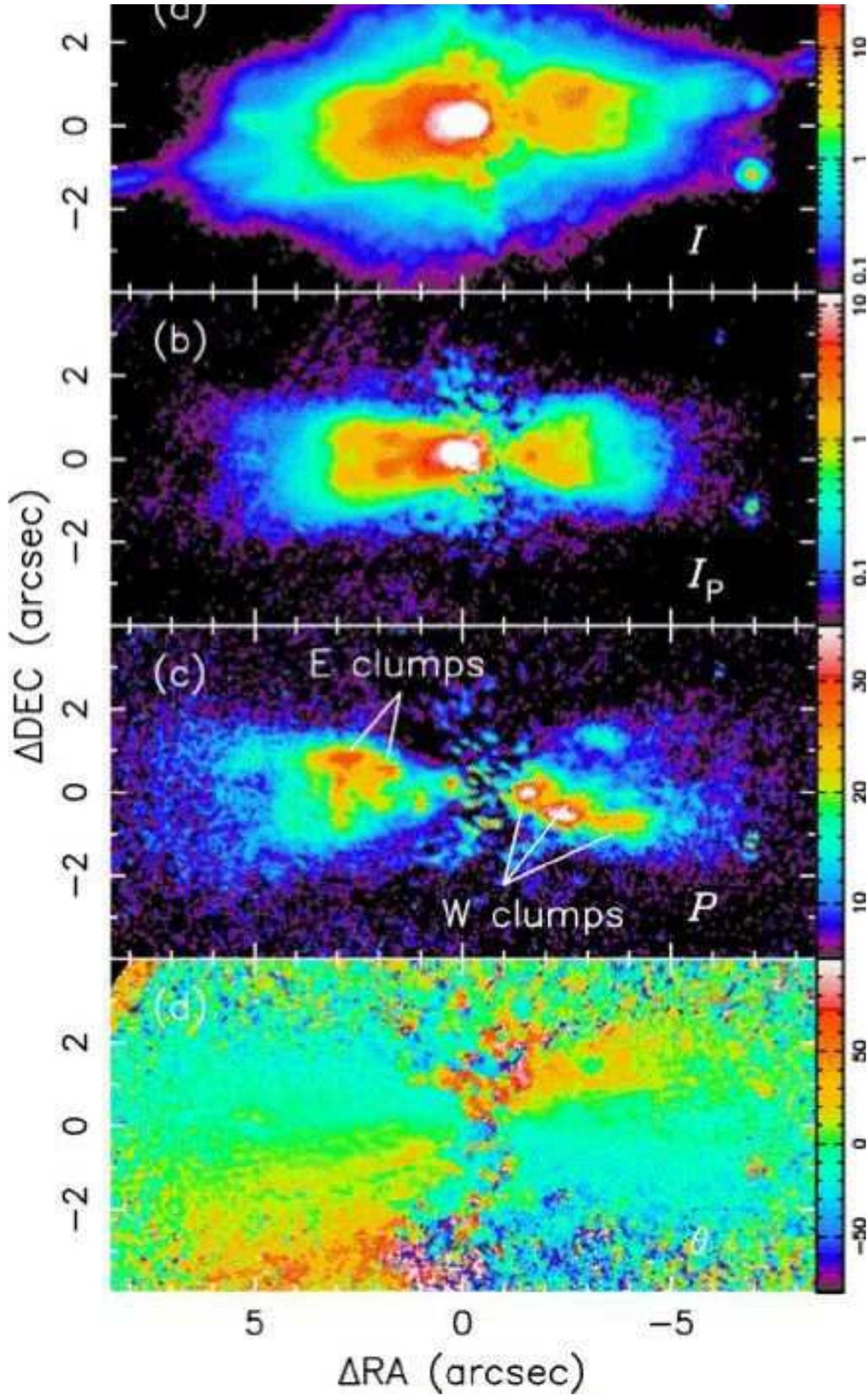}
\caption{\label{04395}%
Imaging polarimetric maps of IRAS 04395$+$3601 in the
(a) total intensity, $I$, 
(b) polarized intensity, {\ipol},
(c) polarization strength, $P$, and
(d) polarization P.A., $\theta$.
The maps are in the standard orientation (north is up, east to the left)
 and centered at the $\theta$ center with tickmarks showing the right
 ascension and declination offsets in arcseconds. 
The wedges indicate the scale of the image tone in data number per
 second (DN s$^{-1}$, which is to be multiplied by PHOTOFNU [see Table 2
 in Paper I] to convert to Jy) for $I$ and {\ipol}, in percent for $P$, and
 degrees east of north in $\theta$ (i.e.\ $\mbox{P.A.~} 0^{\circ}$
 means the polarization vector, which is perpendicular to the scattering
 plane, is oriented in the north-south direction).
The east and west clumps are indicated.}
\end{figure}

IRAS 04395 (Fig.\ \ref{04395}) is a DUPLEX PPN that is on the verge of
becoming a PN (\citealt{westbrook75,cc78}).   
The $I$ map shows its well known bipolar structure elongated in the 
east-west direction due to multiple outflows \citep{trammell02}. 
The lobe interior structure is more clearly revealed in the $I$ map 
than in recent high-resolution ground-based near-IR maps \citep{ufm01}.
PSF subtraction did not improve the quality of the data because
no post-SM3B PSF data with a comparable degree of saturation is
available and the whole PSF structure (especially of the spikes) is
enlarged due to the bright, extended emission core. 

The {\ipol} map shows a general bipolar shape of the nebula superposed
by clumpy structures in the lobes, as has been revealed in the optical
by \citet{trammell02}. 
However, we do not see the tips of the outflows in {\ipol}.
This is consistent with their shock-excited nature \citep{ufm01}.  
The morphology of the nebula changes dramatically in the $P$ map.
Strong polarization is localized in several clumps distributed along the
south edge of the west lobe and the north edge of the east lobe
(respectively marked as W and E clumps in Fig.\ \ref{04395}c).
These clumps are distributed along a curve that delineates an inverse
{\sf S}. 
The $\theta$ map shows general centrosymmetry, and the vector pattern
center and the $I$ peak are roughly coincident. 

The structure of the lobes in {\ipol} and $P$ suggests that there is
more dust on one side of the lobe than the other, given that the lobes 
are optically thin at near-IR.
The regions of enhanced surface brightness in {\ipol} and $P$
spatially correspond to the shortest of the multiple outflows seen in
H$\alpha$ \citep{trammell02}\footnote{The east clumps correspond to the
eastern outflow ``d'' (the northernmost of the three outflows), while
the west clumps coincide with the shortest western outflow (the
southernmost outflow in the west) as presented in Figure 1 of
\citet{trammell02}.}. 
This indicates that these regions of enhanced local density are
associated with the latest outflows that channel recently ejected
matter away from the central region, provided that mass loss proceeds at 
a constant velocity.
In fact, $\theta_{\rm eq}$ ($83^{\circ}$) is aligned with the directions
of these most recent outflows delineated by strong {\ipol} and $P$
($\sim 85^{\circ}$), and is not aligned with the bipolar axis ($94^{\circ}$).  
This implies that the current orientation of the equatorial dust torus
is oblique with respect to the bipolar axis (by about $10^{\circ}$).

The clumpiness in the outflow structures of the enhanced
{\ipol} and $P$ surface brightness suggests mass loss that is episodic.  
\citet{trammell02} have shown multiple ring-like structures in H$\alpha$
and [\ion{S}{2}] emission as well as an arc in H$\alpha$ silhouette.
These structures are in line with the episodic mass loss interpretation.
These clumps, however, could alternatively imply instabilities in the
outflows, as pointed out also by \citet{trammell02}. 

Provided that outflows are associated with episodic mass ejection,
the angular separation between clumps corresponds to the interval of 
mass ejection.
The separations between clumps are $1\farcs1$ and $1\farcs3$ in the west
lobe respectively for the first and second closest clump pairs from
the central star, and $0\farcs9$ in the east lobe.
Using 120 km s$^{-1}$ outflow velocity \citep{trammell02} and the
distance of 1.7 kpc \citep{bujarrabal88} and assuming a linear outflow,
these separations translate to 60 to 90 years. 
Given the uncertainty in the exact geometry of the outflows and of the 
clumps, these values are reasonably close to the value of 40 years
(scaled to our choice of the distance) derived by \citet{trammell02}
based on the separation between the ring-like structures seen in H$\alpha$
and [\ion{S}{2}].

\subsubsection{IRAS 09452$+$1330 (IRC $+$10 216)}

\begin{figure}
\epsscale{.58}
\plotone{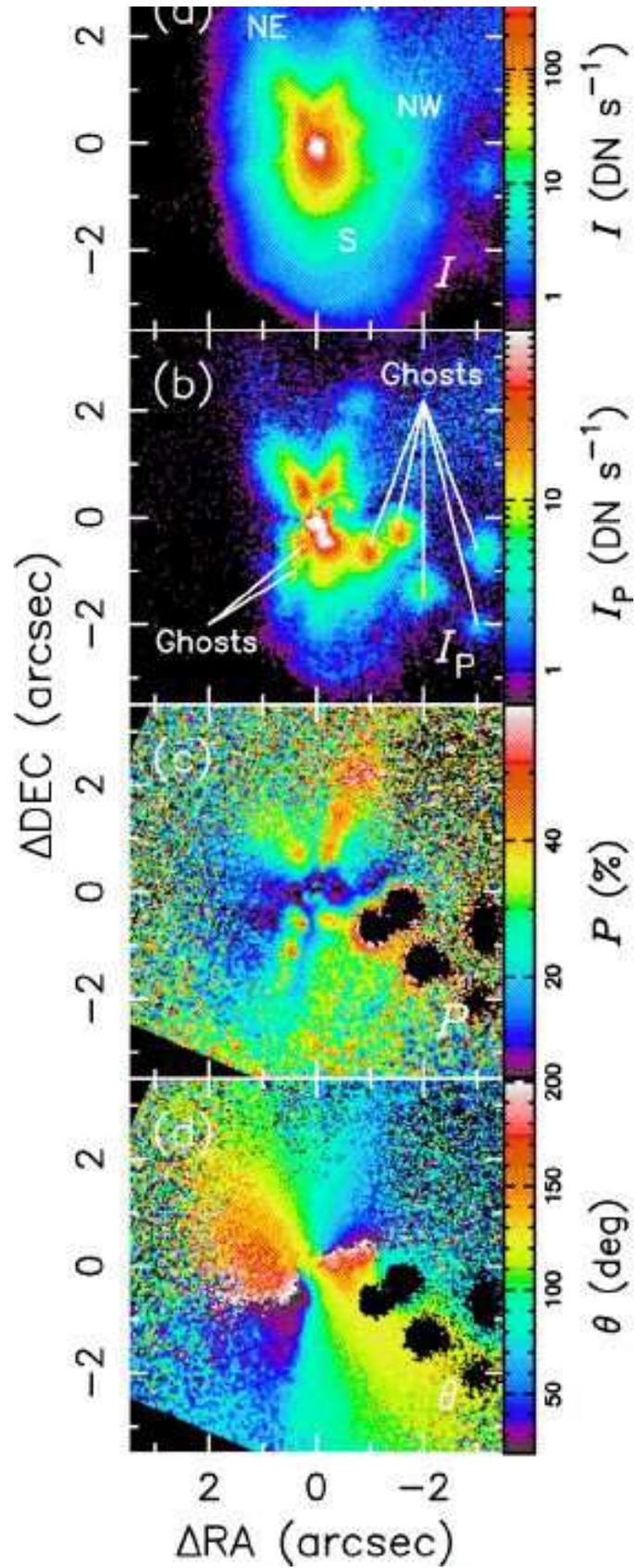}
\caption{\label{09452}%
Same as Fig.\ \ref{04395}, but for IRAS 09452$+$1330.
Each lobe is identified with its designation, NE, N, NW, SE, and SW.} 
\end{figure}

IRAS 09452 (Fig.\ \ref{09452}) is one of the brightest thermal IR sources
in the sky \citep{becklin69}, and hence, one of the best-studied C-rich
AGB star.  
The CSE (of DUPLEX type) is seen as a fragmented bipolar nebula having
four major emission components in the optical and near-IR
\citep{skinner98,oster00,tuthill00,weigelt02,murakawa05}.

The $I$ map shows the main nebulosity that is stretched at a P.A.\ of
$22^{\circ}$. 
We recognize three fragments in the north 
(marked as the ``NE'', ``N'', and ``NW'' in Fig.\ \ref{09452}a
respectively from east to west [$\mbox{P.A.} = 22^{\circ}$,
$-19^{\circ}$, \& $-66^{\circ}$]; \citealt{oster00,tuthill00,weigelt02})
and an arc beyond the south lobe (about $3\arcsec$ away from the central
peak; \citealt{mh00}). 

The {\ipol} map shows the CSE structure more clearly than the $I$ map, 
but it suffers from the polarization ghosts (high {\ipol} regions
located radially with semi-regular intervals in the west and southeast
as indicated in Fig.\ \ref{09452}; \citealt{hines00}).
The northeast and south lobes form the main bipolar nebulae at a P.A.\
of $22^{\circ}$.
However, it is not clear if the north and northwest lobes also have
their counterparts in the southeast because of the ghosts.
\citet{murakawa05} have detected little ($\ale 10\%$) polarization in
the southeast in recent ground-based imaging polarimetry in the H
band and suggested that the north and northwest lobes are caused by
leakage of light through fissures in the west side of the central dust  
torus.
The southeast ghosts are more obvious as in the $P$ map.
These ghosts are also the source of the anomalous polarization vector
pattern in the otherwise centrosymmetric $\theta$ map.

Our aperture polarimetry has yielded $\theta_{\rm eq}$ of $32^{\circ}$.
This is consistent with past polarimetric measurements in the optical
through mid-IR \citep{sz70,dfs71,cd72,ck76,tdg94,serkowski01}.
Therefore, the bipolar lobes are obliquely oriented ($10^{\circ}$ off)
with respect to the equatorial plane. 
Moreover, the southwest lobe appears larger than the northeast lobe.
This is probably due to inclination of the lobes with respect to the
plane of the sky: the angle of $20^{\circ}$ has been determined based on
radiative transfer calculations \citep{skinner98}. 

\subsubsection{IRAS 10131$+$3049 (CIT 6)}

\begin{figure}
\epsscale{.88}
\plotone{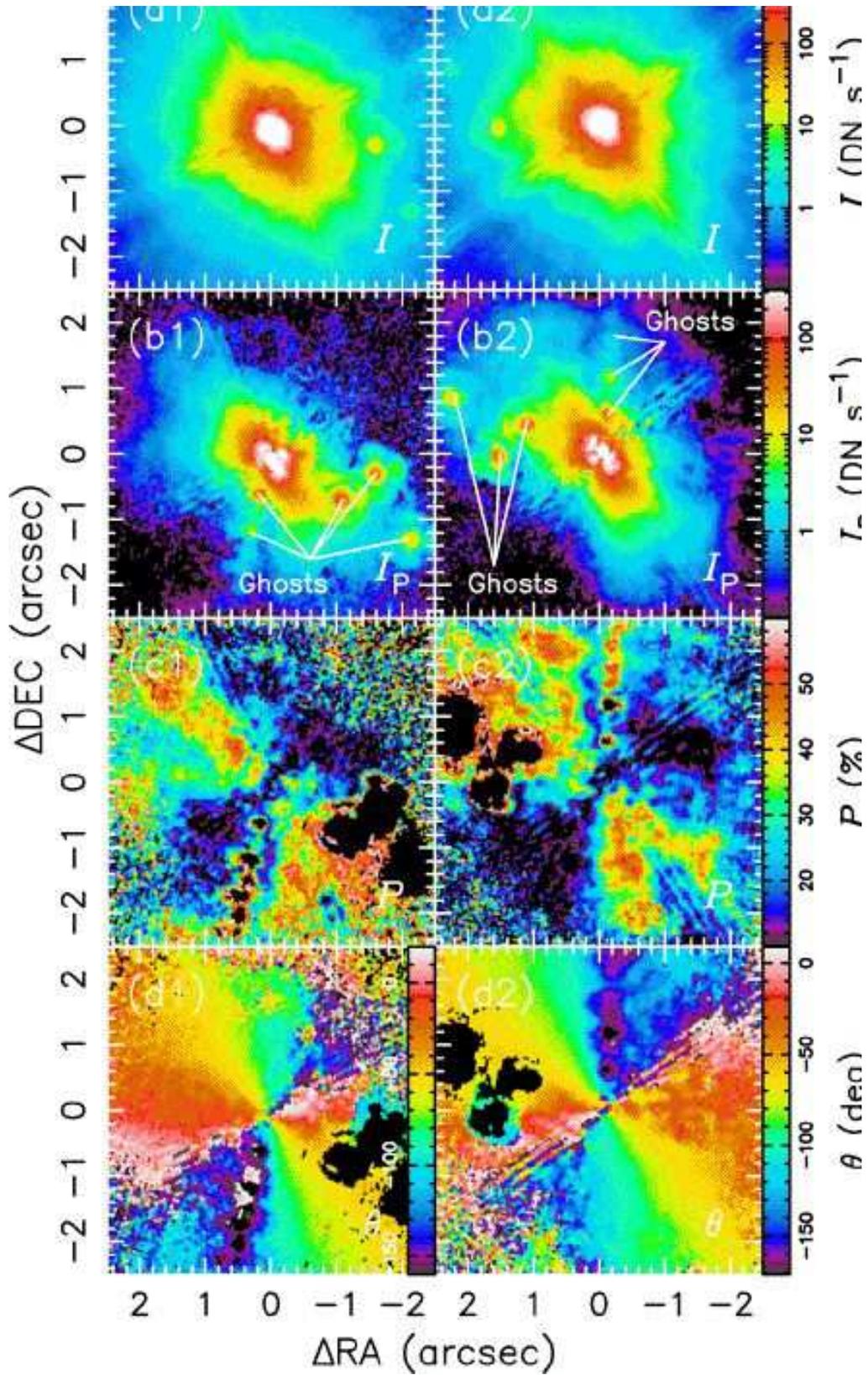}
\caption{\label{10131}%
Same as Fig.\ \ref{04395}, but for the two-epoch observations of IRAS
 10131$+$3049 (Left: 1998 May 5, Right: 1998 Oct 11).  The polarization
 ghosts are indicated.}   
\end{figure}

IRAS 10131 (Fig.\ \ref{10131}) is an IR-bright C-rich extreme AGB
star \citep{ulrich66,cohen79}.
\citet{schmidt02} have studied the CSE structure in detail using a set 
of {\hst} data to show the presence of (1) large IR bipolar lobes
(about 10 times larger than in the optical) in the northeast-southwest
direction and (2) multiple concentric arcs $1\arcsec$ to $4\arcsec$ away
from the central star. 
The two-epoch maps in Fig.\ \ref{10131} show the ghosts affecting all
the quadrants of the field of view since the roll angle of {\hst}
at the two epochs differs by about $180^{\circ}$.
We do not recognize any internal structure in the CSE core in the
{\ipol} and $P$ maps. 

A series of IR polarimetry between 1967 and 1976 showed that $\theta$
varied between $110^{\circ}$ and $180^{\circ}$ \citep{dfs71,k71,kc76}.
Especially, \citet{kc76} reported an IR polarimetric variability on the  
order of several months.
Although \citet{schmidt02} have determined that there is no evidence for
temporal changes in the polarimetric properties of the CSE structure
between the two epochs, there is some $6^{\circ}$ change in 
$\theta_{\rm eq}$.
This may be related to the intrinsic IR polarimetric variability of the
object.   
With respect to the measured $\theta_{\rm eq}$ values, the shell
elongation is not exactly perpendicular ($8^{\circ}$ to $14^{\circ}$
offset). 
If $\theta_{\rm eq}$ varies as $\theta$, it may suggests variations
in the dust density distribution near the equatorial plane on the order 
of months. 

\subsubsection{IRAS 10178$-$5958 (Hen 3-401)}

\begin{figure}
\epsscale{.88}
\plotone{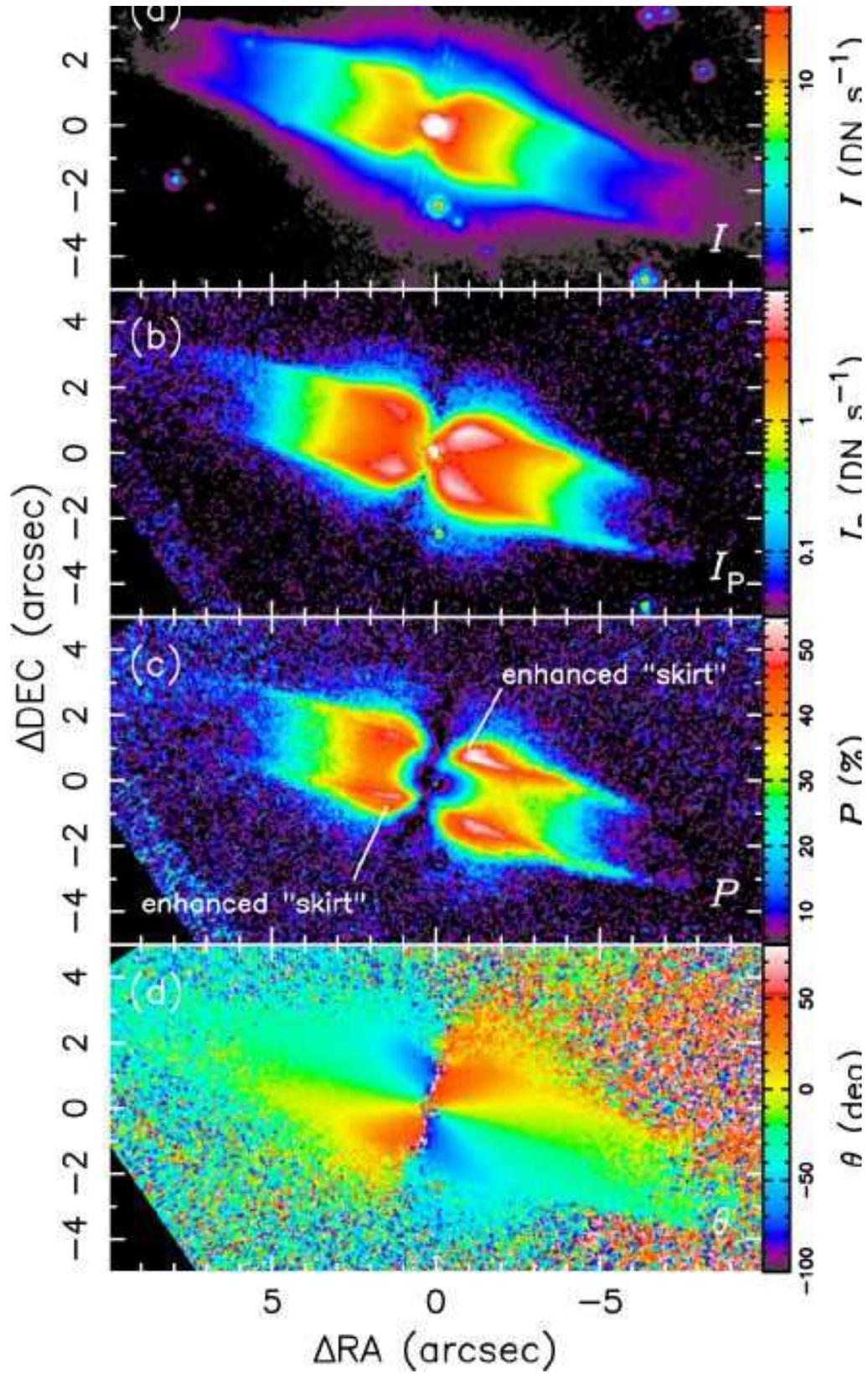}
\caption{\label{10178}%
Same as Fig.\ \ref{04395}, but for IRAS 10178$-$5958.}
\end{figure}

IRAS 10178 (Fig.\ \ref{10178}) is a PPN of DUPLEX type associated with a 
Be post-AGB star.
{\hst} observations in the optical and H$_{2}$ 1-0 S(1) bands have
revealed this object's highly collimated structure
(\citealt{grm99,sahai99}). 
In the $I$ map we see the collimated bipolar lobes that are linearly
elongated into a P.A.\ of 73$^{\circ}$.   
The equatorial region shows a pinched waist due to extinction. 
The inner structure in the IR does not appear as sharply as in the
optical.

The {\ipol} and $P$ maps clearly reveal the physically thin nature of
the walls of the cylindrical lobes.
The dust lane that separates both lobes is more clearly delineated.
Moreover, {\ipol} and $P$ maps show, in the bipolar ``skirt'' region,
that the northwest and southeast walls are more enhanced than the
southwest and northwest walls, respectively.
Furthermore, the southwest and northeast walls appear longer than the
southeast and northwest walls, respectively.
The $\theta$ map indicates a rather aligned distribution of polarization
angles by a shallow gradient of the image tone over the bipolar lobes
(especially away from the central star).
Nevertheless, the $\theta$ map is generally centrosymmetric centered at
the $I$ peak.  

Our measurements yield  $\theta_{\rm eq} = 80^{\circ}$.
The $P$ map shows an elliptically elongated region of weak $P$ ($\ale
20\%$ at a P.A.\ of $78^{\circ}$) superposed on the dust lane.
This weak $P$ region, therefore, is almost orthogonally oriented with
the equatorial plane.
This suggests that this weak $P$ region may represent the region of
diluted polarization due to starlight leaking through the biconical
openings of the dust torus aligned with the equatorial plane.
However, the cylindrical lobes are aligned at a P.A.\ of $73^{\circ}$,
which is slanted by $7^{\circ}$ with respect to the equatorial plane.
In addition, the east end of the northwest wall and the west end of the
southeast wall (the regions marked as ``enhanced skirt'' in Fig.\
\ref{10178}c) seem to be slightly curved towards the central star with
respect to the rest of the walls, which are very much linear.
It appears as if these curved ends of the walls trace the weak
$P$ region.

The polarimetric data thus uncover (1) the point-symmetric distribution 
of scattering matter in the cylindrical lobes (the northwest and
southeast walls are more enhanced) and (2) the misalignment between the
axes of the equatorial plane and the collimated cylindrical lobes.
As a cause for the highly collimated structure of the nebula,
\citet{grm99} on one hand suggested an extreme density contrast within
the dust torus that would preferentially channel outflows into
spatially-confined regions, while \citet{sahai99} on the other proposed
a dissipative hydrodynamic interaction of jets with the surroundings. 
Both scenarios, however, still require additional mechanisms to direct
the outflows/jets along the cylindrical lobes that are not orthogonally
aligned with the equatorial plane. 
Alternatively, the enhancements in the northwest and southeast walls of 
the lobes may be simply due to the illumination effects as suggested
from the shape of the weak $P$ region, since the biconical openings
of the central dust torus (implied by the orientation of the weak $P$
region) are pointed towards these regions of enhanced {\ipol} and $P$.
The pronounced double-shell structure at the east end of the northwest
wall and the west end of the southeast wall may be related to the
misalignment. 

\subsubsection{IRAS 10197$-$5750 (Hen 3-404)}

\begin{figure}
\epsscale{.58}
\plotone{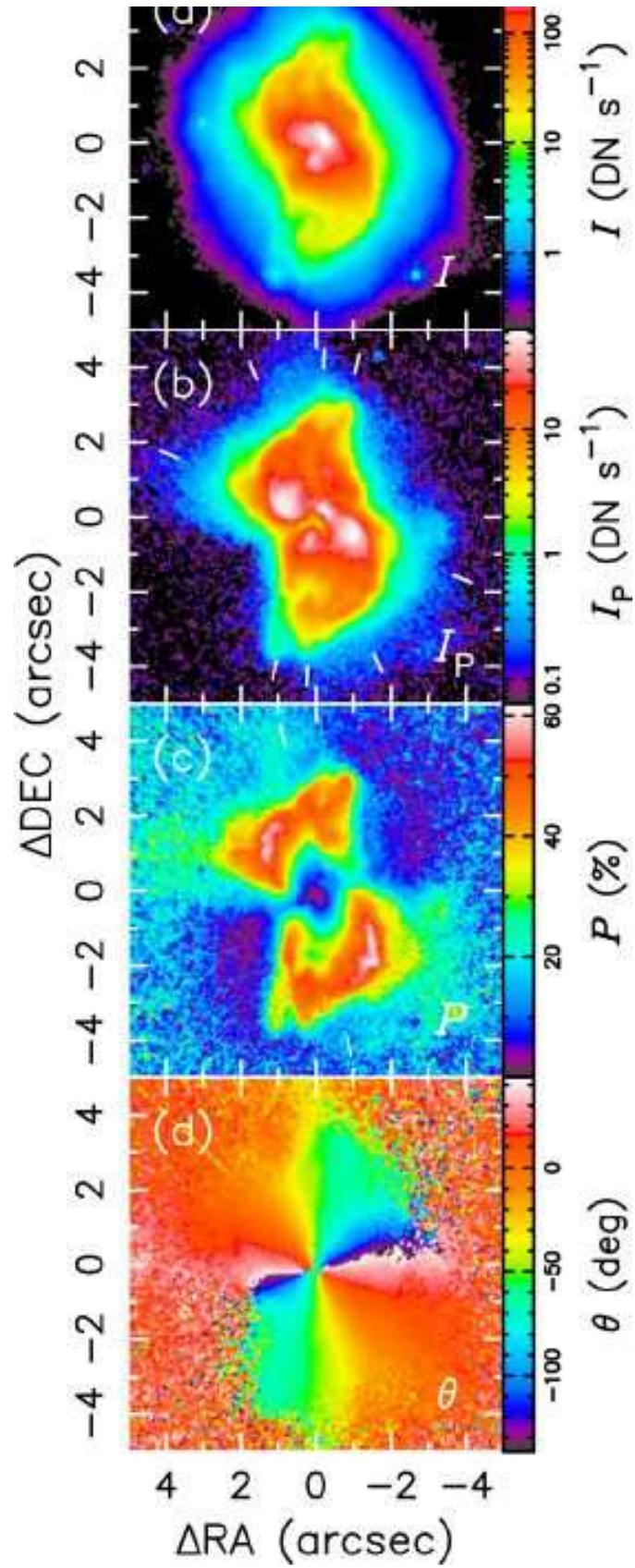}
\caption{\label{10197}%
Same as Fig.\ \ref{04395}, but for IRAS 10197$-$5750.
Radial line segments in the {\ipol} and $P$ maps indicate the directions
 of possible axes of the shell.}
\end{figure}

IRAS 10197 (Fig.\ \ref{10197}) is a bipolar DUPLEX nebula surrounding an
A2I star that is associated with OH maser emission \citep{ahc80}.
The optical reflection nebulosity has been imaged with {\hst} to reveal
its butterfly shape \citep{sahai99b}.
The {\hst} imaging polarimetric data is generally consistent with the
previous imaging polarimetric results presented by \citet{ss95}, but
reveal much more detailed structure.

The $I$ map exhibits some differences in comparison with the optical
map.
Near-IR light delineates a figure {\sf S} rather than a butterfly shape
as in the optical maps, filling the regions of heavy optical extinction
such as the equatorial dust lane and the dark ``spur'' of dust
\citep{sahai99b}.
The {\ipol} and $P$ maps reveal the nebula's point-symmetry better than 
the $I$ map: polarized light is more concentrated on the east edge of
the north lobe and the west edge of the south lobe. 
The distribution of polarized light emphasizes the figure {\sf S}
appearance of the nebula.
The $\theta$ map shows general centrosymmetry.
However, the $I$ peak position does not coincide with the center of
the polarization pattern. 
The location of the central star derived from the polarization pattern
is ($\alpha$, $\delta$)$_{\rm J2000} =  (10^{\rm h}21^{\rm m}33\fs88$,
$-58^{\circ}05^{\prime}47\farcs7)$. 

The $\theta_{\rm eq}$ measurement yields $34^{\circ}$.
Because of the complexity of the nebula, we can define a number of
axes that may characterize the nebula structure as indicated in
the {\ipol} map (Fig.\ \ref{10197}b).
The lobe axis is oriented with a P.A.\ of
(1) $23^{\circ}$ based on the overall structure of the polarized nebula,
(2) $-3^{\circ}$ based on the tips of the figure {\sf S} appearance,
(3) $67^{\circ}$ based on the strongest {\ipol} distribution in the
nebula, and 
(4) $-15^{\circ}$ based on the protrusions along the western edge of the
north lobe and the eastern edge of the south lobe. 
Furthermore, there is a region of weak $P$ ($\ale 20\%$) that is
elliptically elongated at the center of the nebula (similar to what has 
been seen in IRAS 10178). 
This weak $P$ ellipse is oriented with a P.A.\ of $12^{\circ}$
(indicated in the $P$ map, Fig.\ \ref{10197}c). 

None of these axes is exactly perpendicular to the equatorial plane.
If we assume, as in IRAS 10178, that the orientation of the weak $P$
ellipse corresponds to the direction of the biconical openings of the
central dust torus, then the torus is slanted with respect to the
equatorial plane.
In such a case, the central dust torus needs to precess with roughly
$46^{\circ}$ to shape the nebula structure by outflows through the
biconical openings of the torus.
 
\subsubsection{IRAS 12419$-$5414 (Boomerang Nebula)}

\begin{figure}
\epsscale{.58}
\plotone{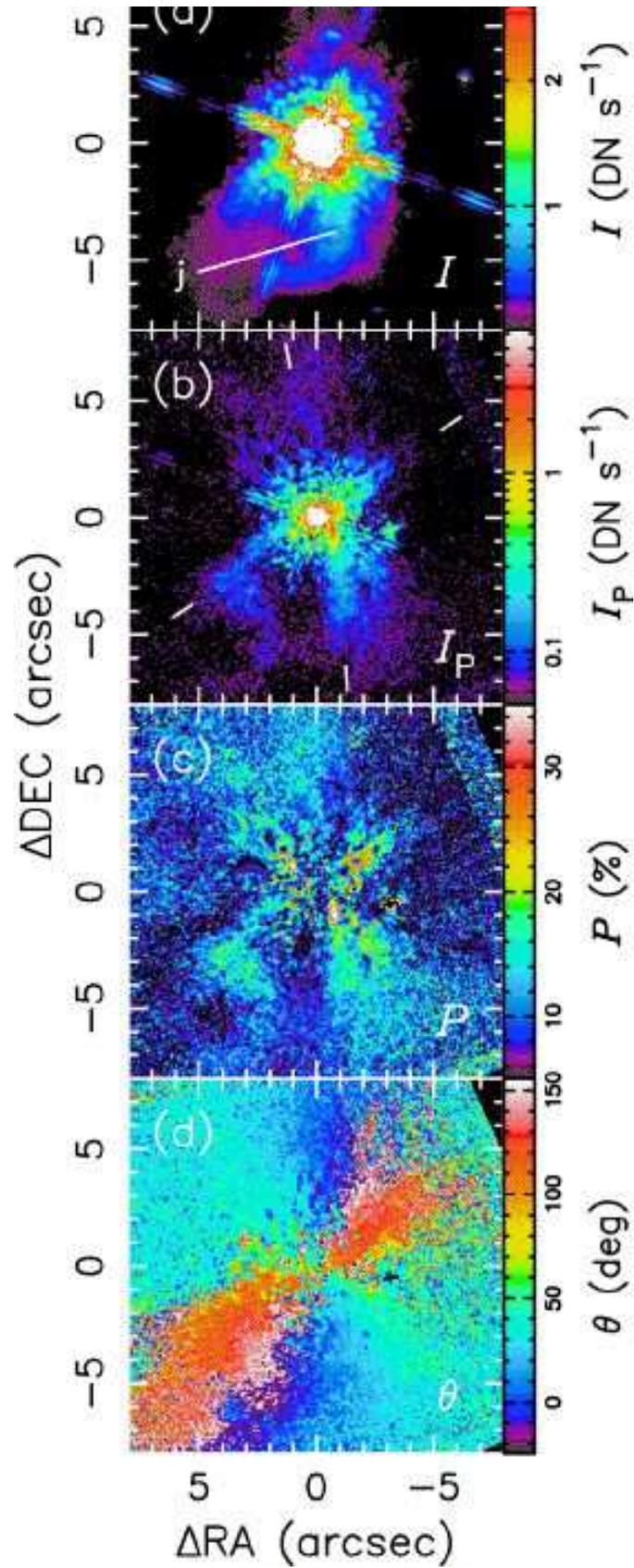}
\caption{\label{12419}%
Same as Fig.\ \ref{04395}, but for IRAS 12419$-$5414.
Radial line segments in the {\ipol} map indicate the directions
 of the bipolar walls.}
\end{figure}

\begin{figure}
\epsscale{.88}
\plotone{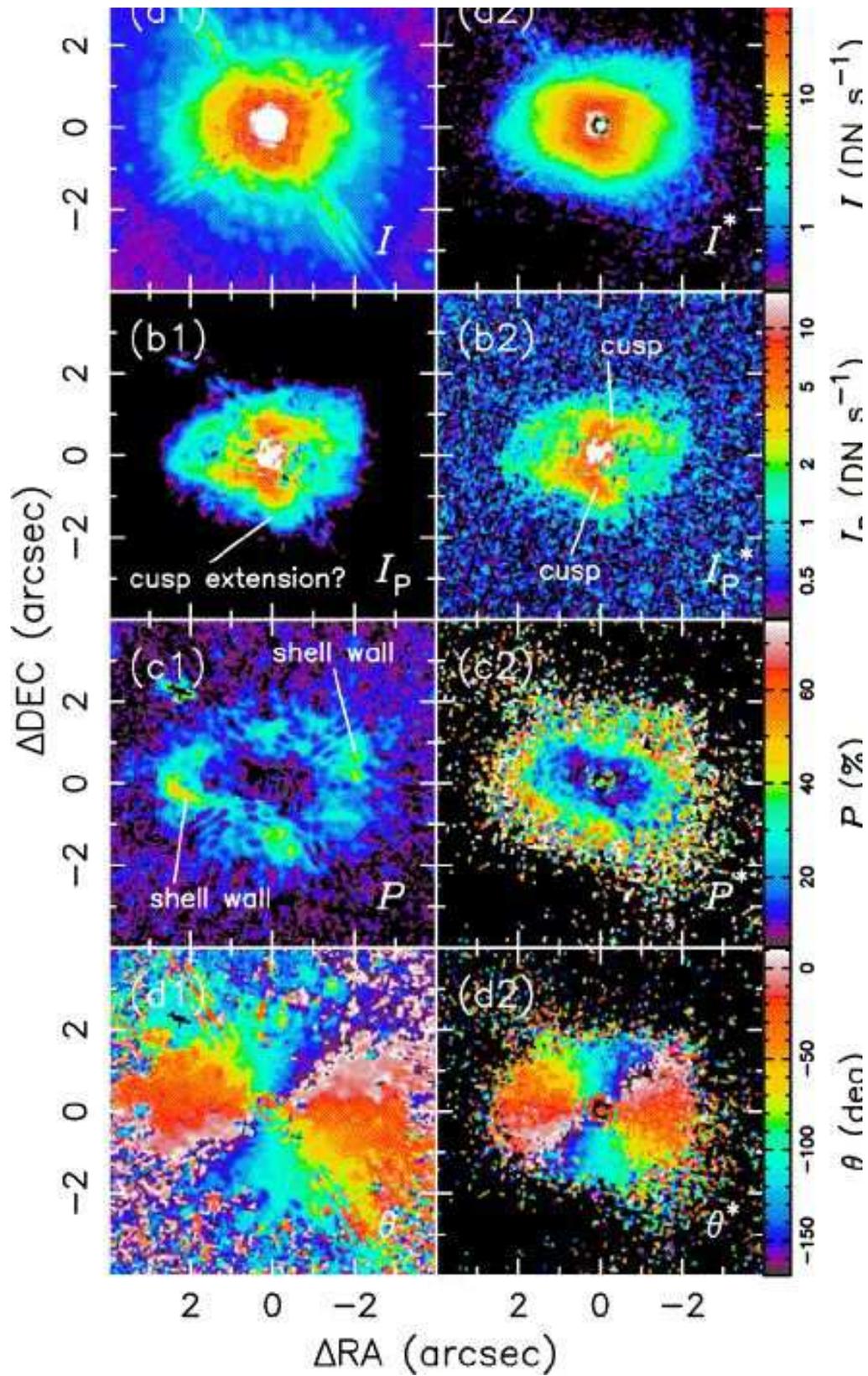}
\caption{\label{16594}%
Same as Fig.\ \ref{04395}, but for IRAS 16594$-$4656 (Left: original, Right: PSF-subtracted).}
\end{figure}

\begin{figure}
\epsscale{.7}
\plotone{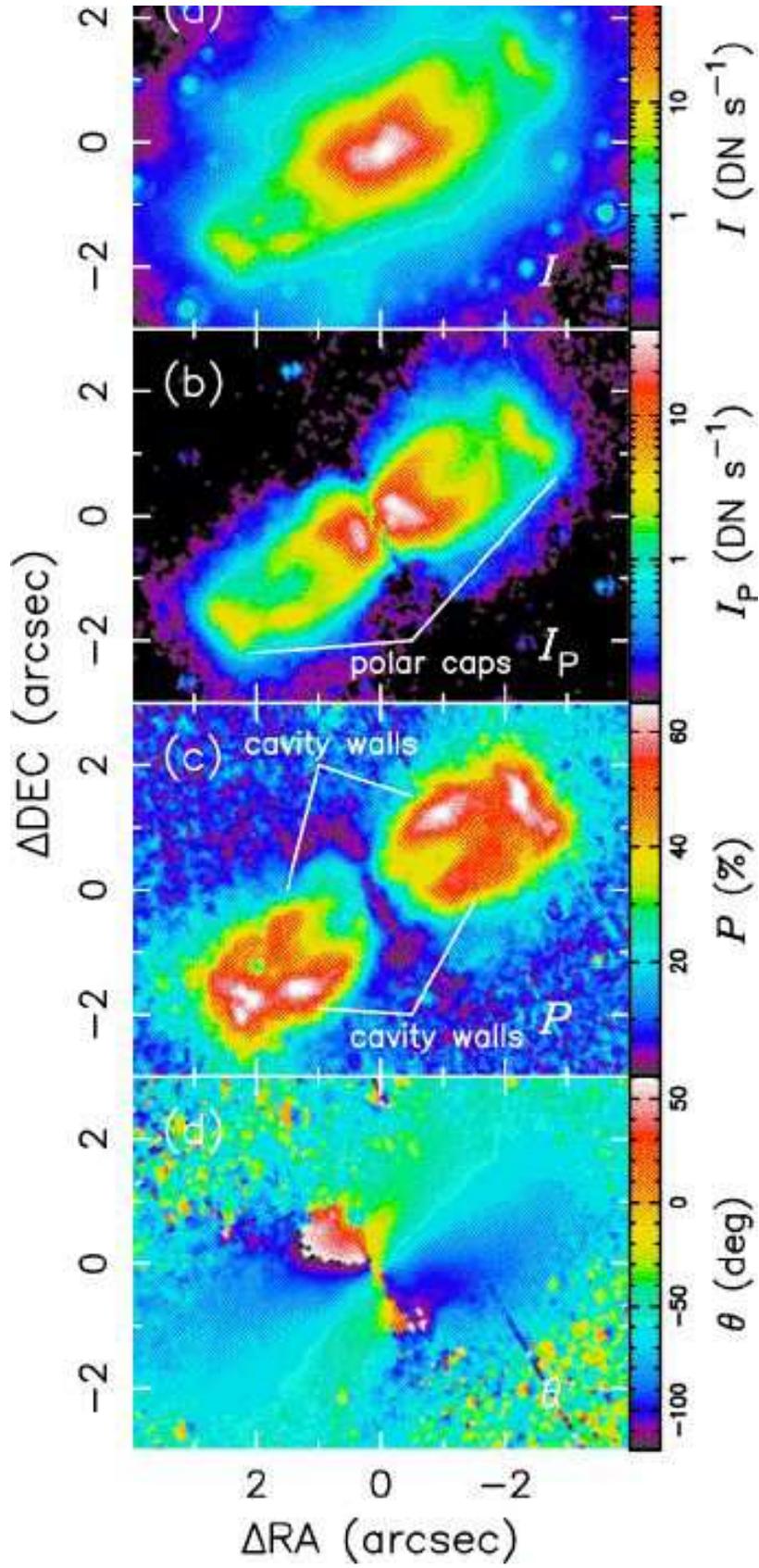}
\caption{\label{17150}%
Same as Fig.\ \ref{04395}, but for IRAS 17150$-$3224.}
\end{figure}

\begin{figure}
\epsscale{.58}
\plotone{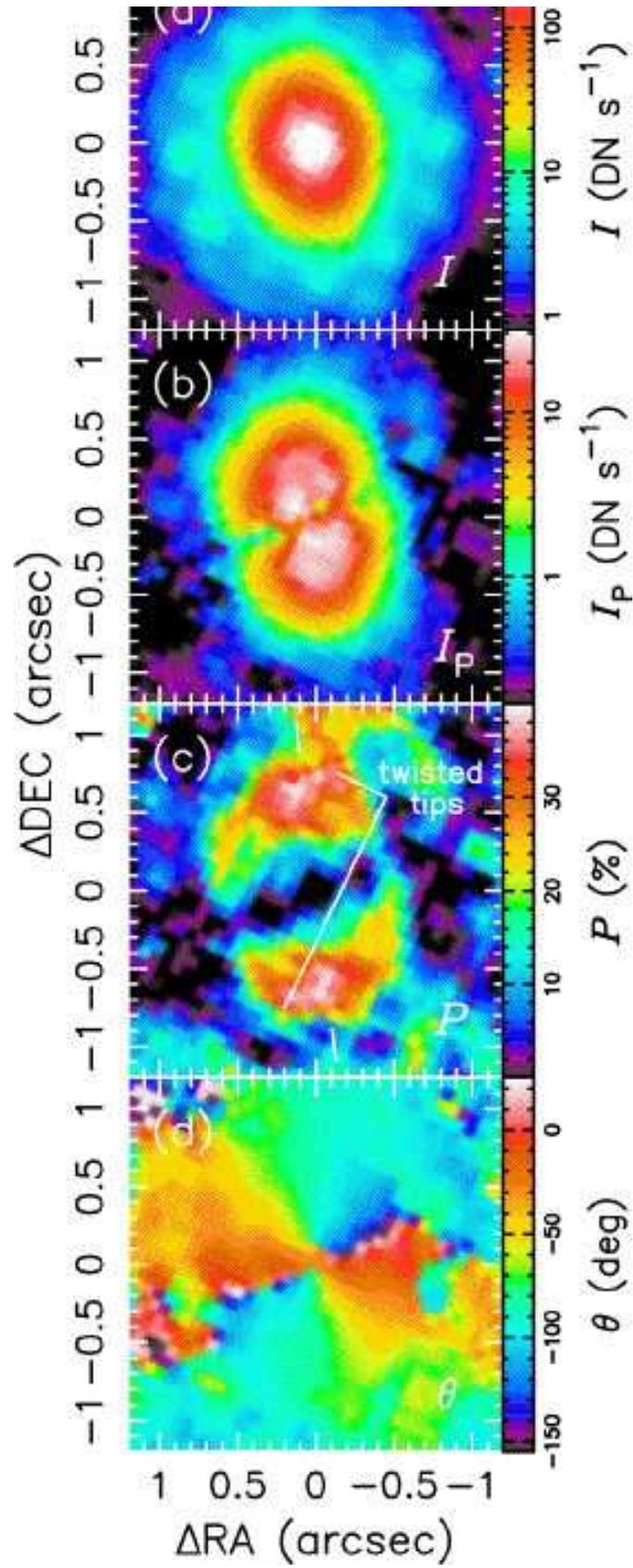}
\caption{\label{17245}%
Same as Fig.\ \ref{04395}, but for IRAS 17245$-$3951.}
\end{figure}

IRAS 12419 is a very large optical bipolar reflection nebula ($55\arcsec
\times 21\arcsec$; \citealt{wg79}), in which very strong optical
polarization has been detected ($45 - 60\%$ in the lobes; \citealt{ts80}). 
The {\hst} data in Fig.\ \ref{12419} show polarization in the central
$8\arcsec \times 8\arcsec$ region of the nebula at $2\micron$. 
In the $I$ map, we see an extremely faint nebulosity in the PSF-dominated
central region, which appears to be shaped in a figure {\sf j} at a
P.A.\ of $0^{\circ}$. 
Although the nebula is too faint to perform PSF subtraction effectively,
the $I$ map shows that the near-IR lobes are elongated along the general
north-south direction (slightly tilted along the northwest-southeast
quadrants). 

The {\ipol} and $P$ maps marginally show the distribution of polarized
light along four ``streaks'' radially emanating from the central
region (P.A.\ of $-170^{\circ}$, $-55^{\circ}$, $125^{\circ}$, and
$10^{\circ}$ as indicated in Fig.\ \ref{12419}b).
These streaks seem to delineate the edges of the bipolar lobes.
The $\theta$ map pattern appears generally centrosymmetric.
Although the degree of polarization is similar in the near-IR and
optical around the central star ($12\pm8\%$ and $15\%$, respectively),
the polarization angle is not the same ($147^{\circ}$ and $80^{\circ}$,
respectively).   
This is probably caused by the matter in the equatorial region that is 
optically thick in the optical but optically thin in the near-IR.

\subsubsection{IRAS 16594$-$4656}

IRAS 16594 (Fig.\ \ref{16594}) is an evolved post-AGB star associated
with a multi-funneled optical reflection nebula, in which
shocks start to shape the shell structure \citep{hrivnak99,vds03}. 
The imaging-polarimetric data of this object have been presented by
\citet{su03} and us (Paper I). 
Since this object presents an interesting case of a rather optically
thin DUPLEX PPN, we briefly re-describe the structure of the nebula.
This object possesses 
(1) an elliptical nebula elongated in the east-west direction (P.A.\ of
$82^{\circ}$) with an additional protrusion towards northwest (P.A.\ of
$-60^{\circ}$) and  
(2) the bipolar cusps in {\ipol} that delineate the innermost
edges of the equatorial density enhancement in the shell.
The cusps are spatially coincident with H$_{2}$ emission
\citep{hrivnak04}, which shows enhancements at the edges of the
equatorial torus in an otherwise elliptically elongated shell (aligned
with one of the optical funnels).  
$P$ is slightly enhanced on the south edge in the east lobe and on the
north edge in the west lobe. 
The vector pattern is generally centrosymmetric.

$\theta_{\rm eq}$ is found to be $22^{\circ}$, which is unrelated
to any of the observed structure.
While IRAS 16594 is of DUPLEX type based on its shell structure, the fact
that the central star is visible in the optical and near-IR suggests
that the optical depth in the equatorial region is near the lower limit
for DUPLEX (about unity).
Moreover, the two-peaked mid-IR core structure found in the equatorial
region \citep{gh04} also indicates rather low-density enhancement.  
Thus, our $\theta_{\rm eq}$ measurement may have been affected by
the bright PSF.

\subsubsection{IRAS 17150$-$3224}

IRAS 17150 (Fig.\ \ref{17150}) is an archetypical DUPLEX PPN with a
highly equatorially enhanced shell (\citealt{meixner02} and references 
therein).
The $I$ map reveals the equatorial region of the nebula, which is very
much obscured in the optical \citep{kwok98,ueta00}.
{\hst} images in near-IR bands have been presented by \citet{su03}. 
From the polarization pattern, we find that the position of the central
illumination source is ($\alpha$, $\delta$)$_{\rm J2000} = 
(10^{\rm h}17^{\rm m}19\fs81$, $-32^{\circ}27^{\prime}21\farcs2)$, which
is $0\farcs14\pm0\farcs07$ off from the central $I$ peak.

The {\ipol} and $P$ maps display the bipolar structure of the nebula, in
which the hollowness of the lobes is highlighted by the polarized light
delineating the edges and polar caps.
In addition, $P$ is even more strengthened in the north wall of the
northwest lobe and in the south wall of the southeast lobe.
Thus, there is a preferential distribution of scattering matter in a
figure {\sf S}. 
The measured $\theta_{\rm eq}$ is 120$^{\circ}$, which is consistent with
the bipolar axis at a P.A.\ of $122^{\circ}$.
 
\subsubsection{IRAS 17245$-$3951}

IRAS 17245 (Fig.\ \ref{17245}) is a less studied post-AGB star
\citep{hrivnak99}, whose less evolved nature has been suggested by the
star's association with OH maser emission \citep{sevenster02}.
The polarization maps reveal the bipolar nature of the nebula
that does not possess an apparent bipolar nebulosity in the total
intensity maps \citep{gledhill01,su03}.   
While the main bipolar lobes are oriented along the north-northeast to
south-southwest direction, the tail end of the nebula tips appears to be
twisted towards opposite directions: the north tip to northwest and the
south tip to southeast.  
This tendency is seen more clearly in the $P$ map, in which 
the strongly polarized tips are found at a P.A.\ of $7^{\circ}$.
The bipolar axis of the overall nebula elongation is oriented at a P.A.\
of $15^{\circ}$, while $\theta_{\rm eq}$ is found to be 21$^{\circ}$.
Thus, there is $6^{\circ}$ misalignment between the bipolar axis and the
orientation of the equatorial plane. 

\newpage
\subsubsection{IRAS 17423$-$1755 (Hen 3-1475)}

\begin{figure}
\epsscale{.58}
\plotone{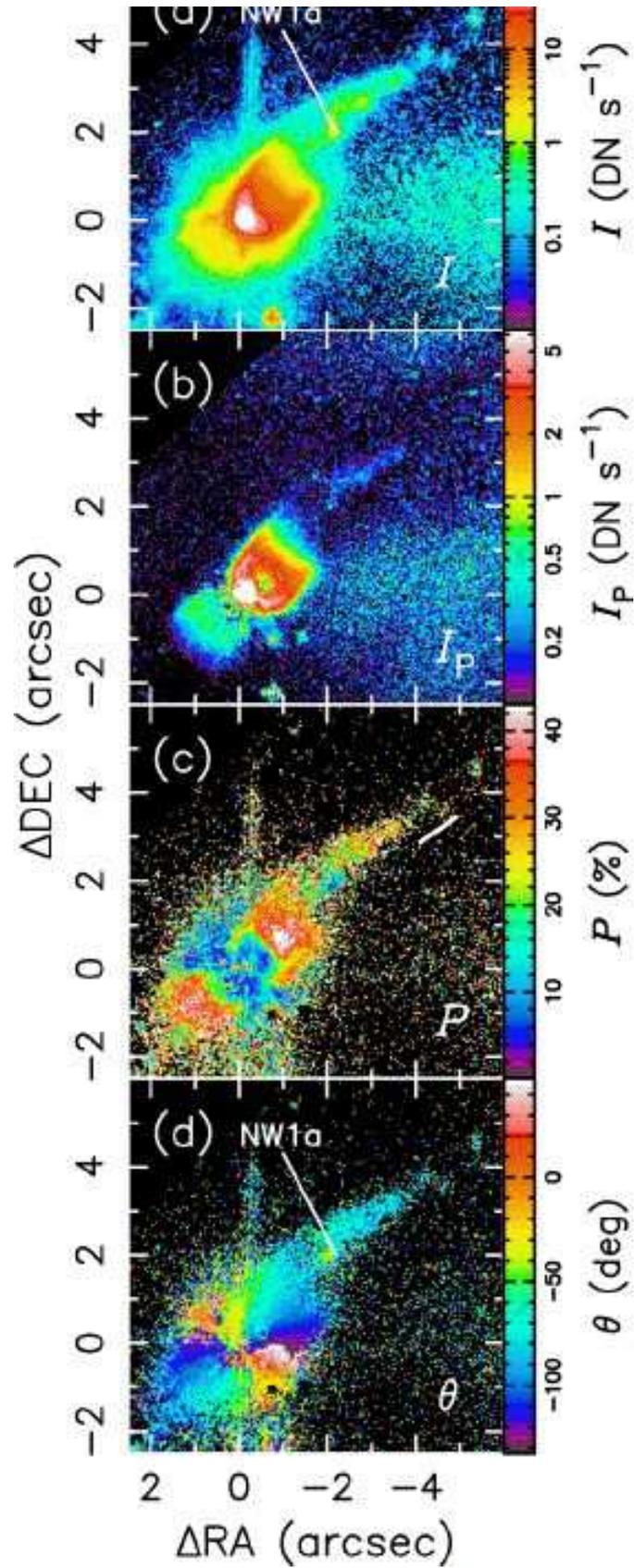}
\caption{\label{17423}%
Same as Fig.\ \ref{04395}, but for IRAS 17423$-$1755.
The position of the NW1a knot \citep{riera03} is shown in the $I$ map.
The radial line segment in the $P$ map indicates the orientation of the
 strong $P$ regions in the bipolar lobes.}
\end{figure}

\begin{figure}
\epsscale{.58}
\plotone{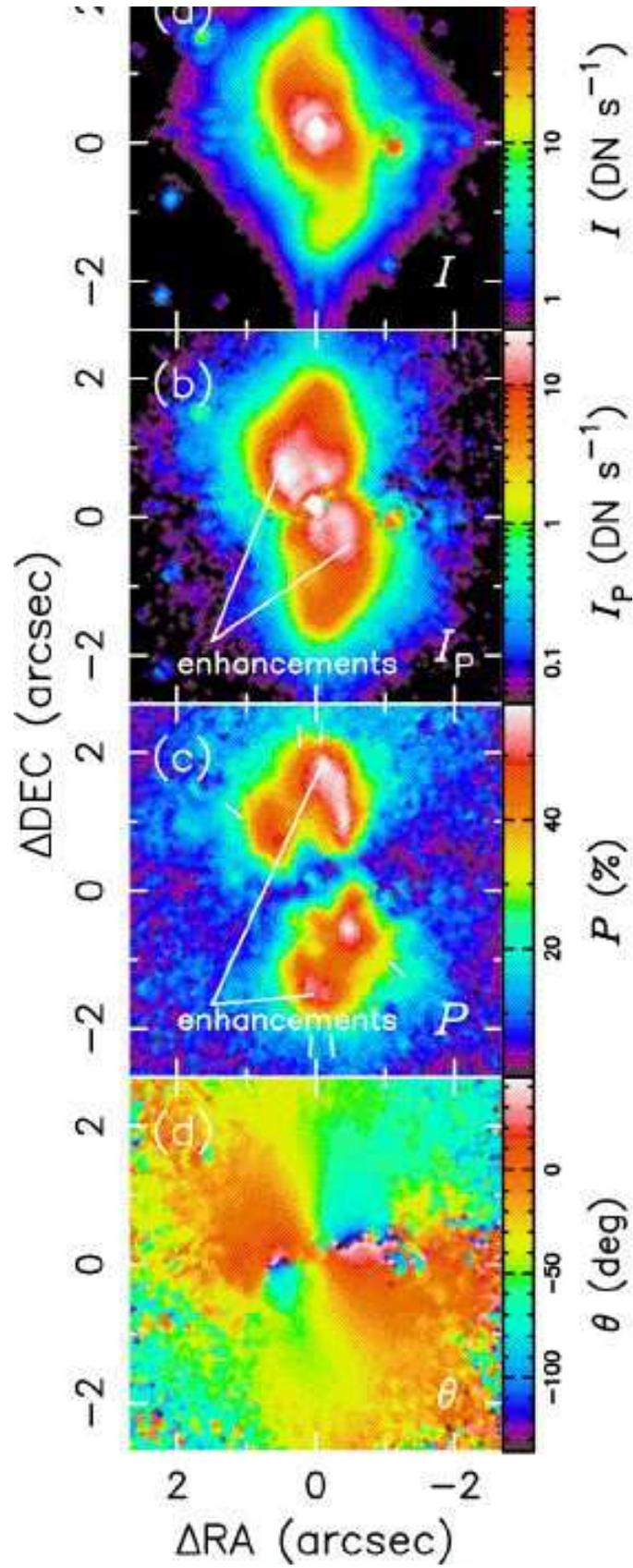}
\caption{\label{17441}%
Same as Fig.\ \ref{04395}, but for IRAS 17441$-$2411.
Radial line segments in the $P$ map indicate the directions
 of possible axes of the shell.}
\end{figure}

\begin{figure}
\epsscale{.88}
\plotone{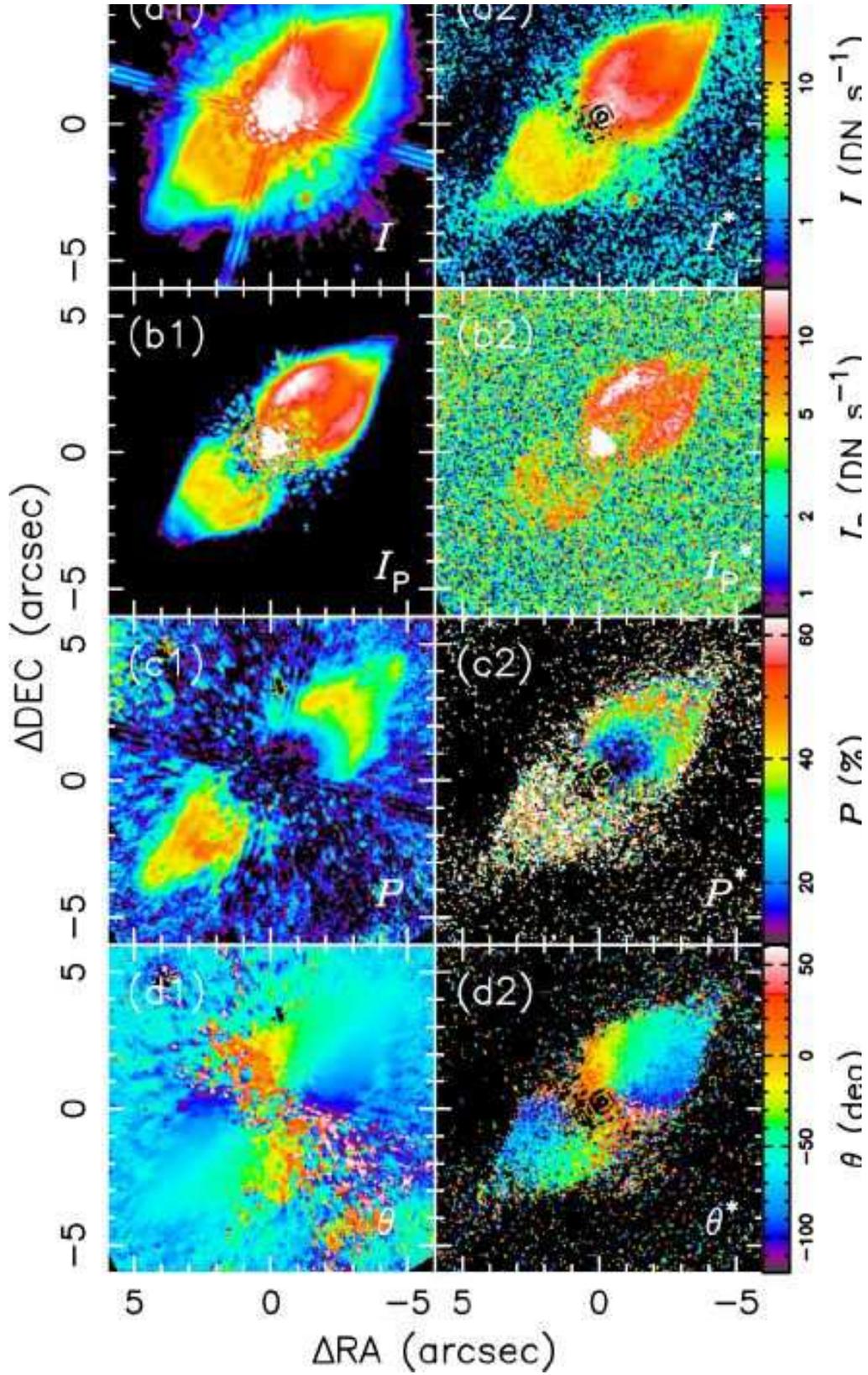}
\caption{\label{19343}%
Same as Fig.\ \ref{16594}, but for IRAS 19343$+$2926 (Left: original,
 Right: PSF-subtracted).}
\end{figure}

\begin{figure}
\epsscale{.88}
\plotone{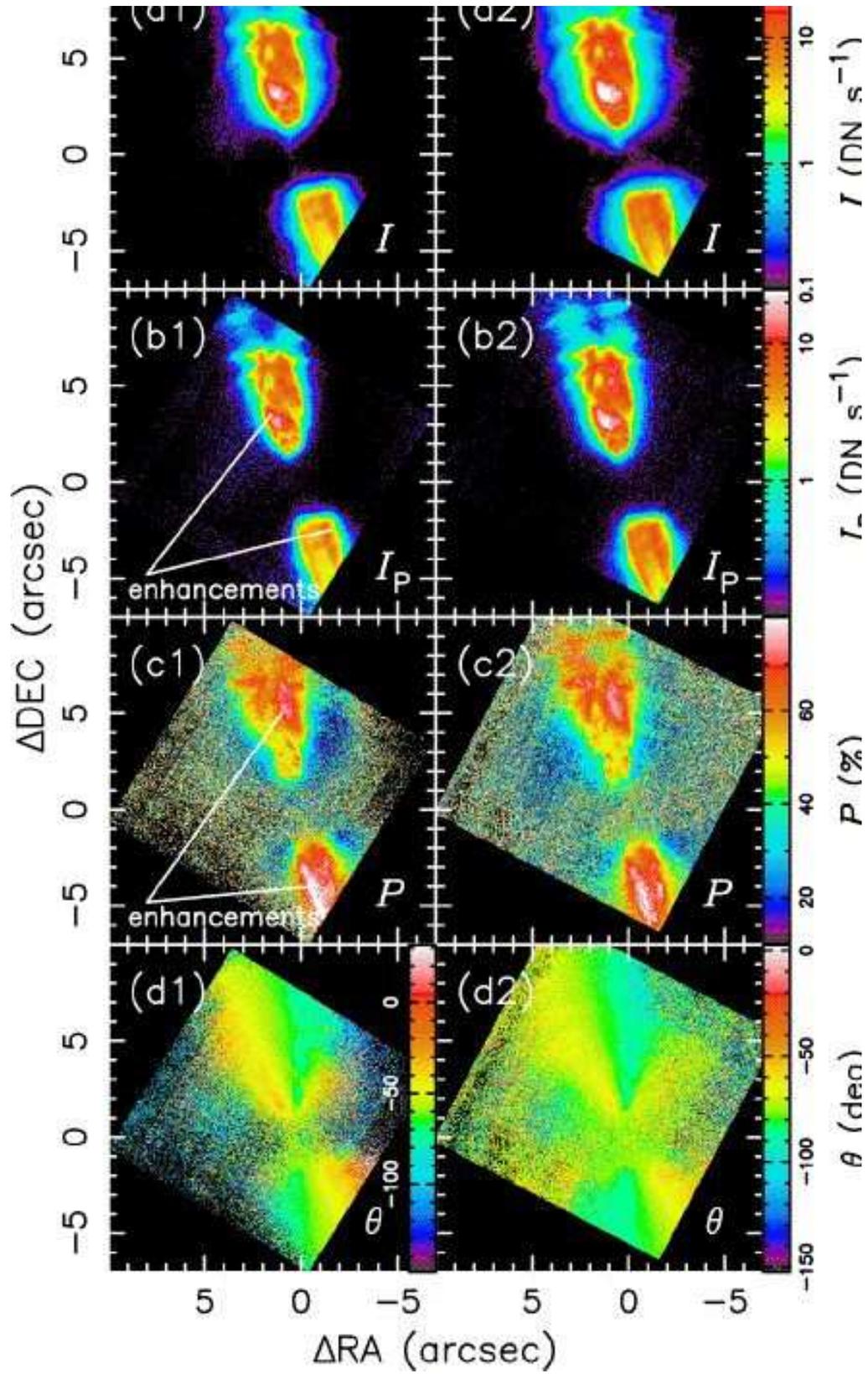}
\caption{\label{2688s}%
Same as Fig.\ \ref{04395}, but for the two-epoch observations of AFGL 2688
 in the NIC1 band (POL-S; Left: 1998 Apr 8, Right: 2002 Oct 6).}
\end{figure}

IRAS 17423 (Fig.\ \ref{17423}) is a highly collimated bipolar PPN
(of DUPLEX) associated with high velocity outflows (e.g.\
\citealt{ss01,riera03}). 
The present data cover only part of the brighter northern lobe in the
NIC1 band. 
The $I$ map mainly shows the northwest lobe of the central
hourglass-shaped bipolar structure that is inclined with respect to the
plane of the sky.  
The extended knots are seen in the $I$ map (up to the NW2 knot;
\citealt{riera03}). 
The {\ipol} map reveals the physically thin walls of the bipolar lobes more
clearly. 

Strong polarization is found close to the west side of the north lobe
and to the east side of the south lobe (in the central ``hourglass''
part). 
These strong $P$ regions are found at a P.A.\ of $-54^{\circ}$ (indicated
in Fig.\ \ref{17423}c). 
This angle is nearly orthogonal to the equatorial plane 
($\theta_{\rm eq} = 128^{\circ}$).
There is $5^{\circ}$ misalignment between the equatorial plane and the
general orientation of the bipolar lobes (at a P.A.\ of $133^{\circ}$). 
Thus, these strong $P$ regions can be interpreted as the particular
locations in the walls of the bipolar lobes that are illuminated by
the central star through the biconical openings of the central dust
torus, which is not orthogonal to the axis of the bipolar lobes.

\subsubsection{IRAS 17441$-$2411}

IRAS 17441 (Fig.\ \ref{17441}) is a G0I post-AGB star associated with an
optical nebula of DUPLEX type \citep{su98,ueta00}. 
The figure {\sf S} shape of the nebula is more pronounced in the near-IR
than in the optical, as seen in the $I$ map.
The $I$ peak coincides with the polarization pattern center, and thus,
represents the central source obscured in the optical. 

As in IRAS 10197, a number of axes can be defined based on various
morphological/polarimetric characteristics.
The tips of the figure {\sf S} shell point to a P.A.\ of
$-2^{\circ}$, while the inner region of the nebula, where {\ipol} is the
strongest, is elongated along a P.A.\ of $46^{\circ}$.
The $P$ map reveals the walls of the bipolar lobes, which are generally
oriented towards a P.A.\ of $6^{\circ}$.
However, the central weak $P$ region ($\ale 20\%$) appears elongated
along a P.A.\ of $20^{\circ}$. 
Aperture polarimetry determines $\theta_{\rm eq}$ of $28^{\circ}$, which
agrees with none of the characteristic P.A.s described above.

\citet{oppen05} have constructed a detailed model of this object using
evacuated bipolar lobes with an equatorial disk by fitting their
polarization data from optical to near-IR.
Their model is quite successful in reproducing the highly polarized
nature of the shell as well as the axisymmetric overall geometry of the
object. 
However, their model does not address the point-symmetric aspect of the
shell structure.
The polarization structure is peculiar in IRAS 17441 because {\ipol} is
enhanced in the east edge of the north lobe and the west edge of the
south lobe, but $P$ is enhanced in the opposite side in each lobe.

\subsubsection{IRAS 19343$+$2926 (M 1-92)}

IRAS 19343 (Fig.\ \ref{19343}) is an O-rich PN of DUPLEX type, which
has been extensively studied
(\citealt{trammell96,bujarrabal98a,bujarrabal98b}, and references
therein). 
The bipolar lobes that are separated by the dust lane are seen in
the $I$ maps as have been seen in the previous observations in atomic
and H$_{2}$ line emission \citep{bujarrabal98b} as well as in CO
\citep{bujarrabal94,bujarrabal98a}.
We do not see, however, the chains of spots in the bipolar cavities that
have been identified as shock-excited clumps of material \citep{bujarrabal98b}. 
Their non-detection in scattered light is consistent with their proposed
shock-excited nature and suggests very little presence of scattering
matter around the shocked gas.

The overall polarization characteristics are generally consistent with 
the previous optical imaging polarimetry \citep{schmidt78}.
The present data show that 
(1) {\ipol} is concentrated in the walls of the bipolar lobes, revealing
the hollow nature of the lobes, and 
(2) strong $P$ is found along the north edge of the northwest lobe and
the south edge of the southeast lobe of the nebula that is generally 
centrosymmetric. 
The nebula shape defines the symmetric axis (P.A.\ of $131^{\circ}$),
along which CO outflow velocity is found to have an almost constant
latitudinal gradient \citep{bujarrabal98a}.
This axis is, however, not aligned with $\theta_{\rm eq}$ of
151$^{\circ}$. 
As in other objects, we find some $P$ enhancements in the nebula
along the opposite edges in each lobe.

\subsubsection{AFGL 2688 (Egg Nebula)}

\begin{figure}
\epsscale{.88}
\plotone{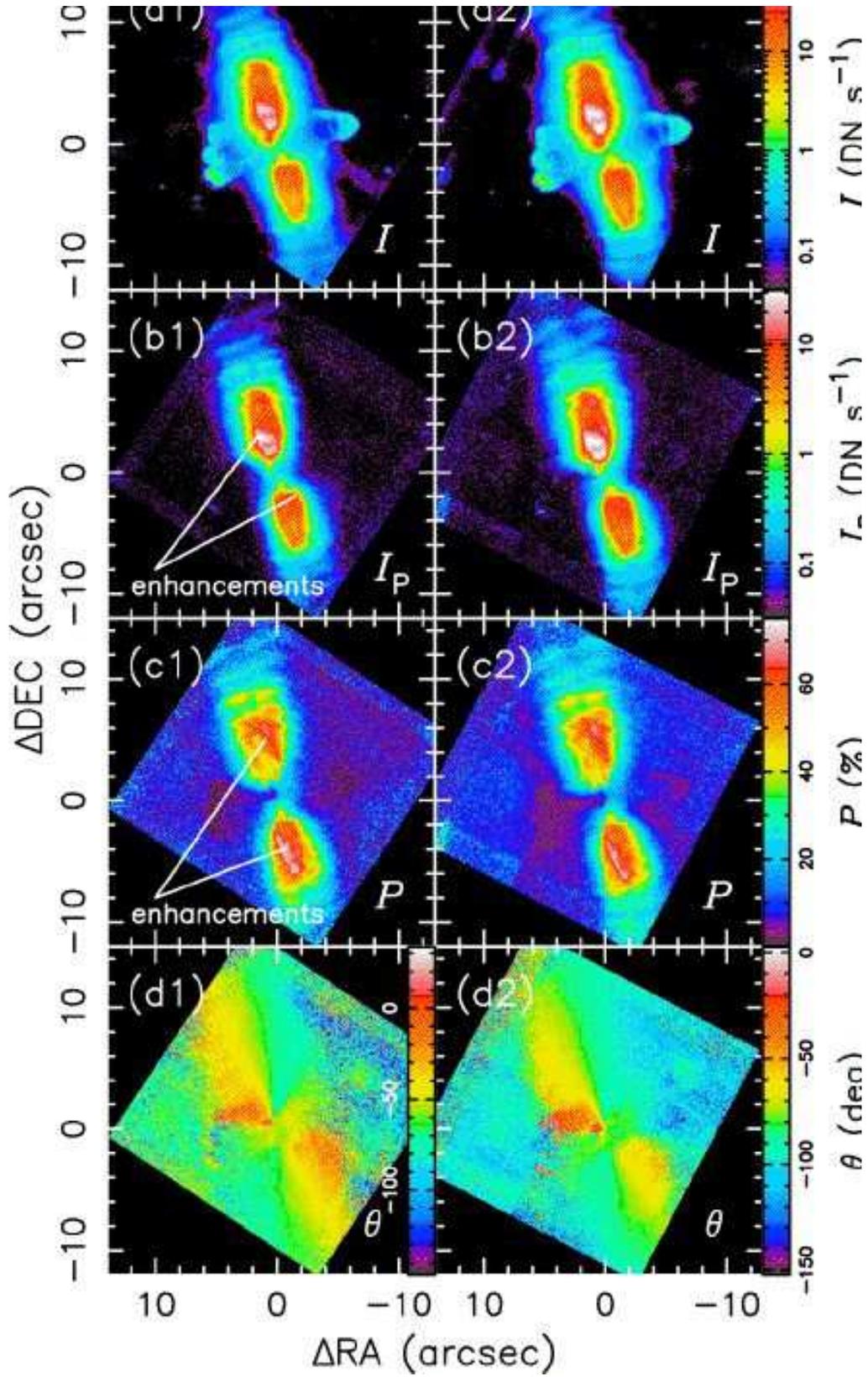}
\caption{\label{2688l}%
Same as Fig.\ \ref{2688s}, but for the NIC2 band (POL-L) observations.}
\end{figure}

AFGL 2688 (Figs.\ \ref{2688s} and \ref{2688l}) is a DUPLEX PPN that is
considered one of the archetypes \citep{ney75,sahai98}, and
its structure has recently been reviewed in detail by \citet{goto02}.
The $I$ maps show the object's well known bipolar lobes plus the
concentric arc structure and the ``spindle'' structure along the 
dust lane between the lobes (seen only in the NIC2 band; Fig.\
\ref{2688l}a).  
The {\ipol} maps display the point-symmetric nature of the lobes with
the brightness enhancement appearing near the southeast corner of the
north lobe and near the northwest corner of the south lobe.
It is remarkable that the spindle structure has almost zero
polarization. 
This is because the spindle structure is largely in H$_{2}$ emission
\citep{cox97,kastner01}.   

In the $P$ maps, strong polarization arises near the northwest corner of 
the north lobe and near the southeast corner of the south lobe. 
The equatorial region of the nebula is optically thick even to near-IR
light, but optically thin lobes show a generally centrosymmetric pattern
in the $\theta$ map from which the position of the illumination source
is determined \citep{weintraub00}.
$\theta_{\rm eq}$ ($17^{\circ}$ to $19^{\circ}$ P.A.) is slightly
misaligned with respect to the axis of the nebula (P.A.\ of
$12^{\circ}$; \citealt{weintraub00,goto02}). 
Proper-motion measurements of the shell structure using the two-epoch
NICMOS data have yielded the distance of 420 pc to the object
\citep{ueta05b}. 

\subsubsection{IRAS 07134$+$1005, 06530$-$0213, 04296$+$3429, and (z)02229$+$6208}

Imaging-polarimetric characteristics of these SOLE objects have already 
been discussed in detail (Paper I).
Here, we briefly summarize our findings.

IRAS 07134 (Fig.\ 3 of Paper I) is a C-rich F5Iab post-AGB star
\citep{hrivnak89} associated with an elliptical nebula 
\citep{meixner97,dayal98,jura00,ueta00,kwok02}.
The shell is a slightly prolate spheroid with an inner cavity and a
built-in equatorial enhancement, and hence, is not simply ``toroidal'' 
as has been interpreted by previous mid-IR observations
\citep{meixner97,dayal98,jura00,kwok02}.   
In addition, the {\ipol} map confirmed that there is more matter on the
east side of the nebula than the west \citep{meixner04}. 

IRAS 06530 (Fig.\ 4 of Paper I) is a C-rich F5Ib post-AGB star
\citep{hrivnak03} with an elliptical reflection nebula \citep{ueta00}.
The nebula is a nearly edge-on prolate spheroid with a possible inner
cavity.   
The nebula's density structure appears such that the equatorial region
of the cavity is filled with some matter, which manifests itself as the
density-enhanced bipolar cusp-like structure. 
However, the total amount of dust along the equator is still not large
enough to cause self-extinction within the shell (but close to the upper
limit for a SOLE nebula). 

IRAS 04296 (Fig.\ 5 of Paper I) is a C-rich G0Ia post-AGB star
\citep{hrivnak95} that is associated with an {\sf X}-shaped optical
elongation \citep{sahai99a,ueta00}. 
While \citet{sahai99a} suggested that these elongations represent
a pair of bipolar jets and a collimating disk, we interpreted that the  
shell had two, most likely hollow, spheroids: one (P.A.\ of
$26^{\circ}$) is nearly edge-on and the other (P.A.\ of $99^{\circ}$) is
inclined with respect to the plane of the sky.
The two-spheroid interpretation was favored by a recent dust scattering
modeling of the object \citep{oppen05}. 

IRAS 02229 (Fig.\ 6 of Paper I) is a C-rich G8$-$K0Ia post-AGB star
\citep{hk99} surrounded by an elliptical nebula of dust-scattered
starlight \citep{ueta00} and dust emission \citep{kwok02}.
This object is the most peculiar among these four SOLE nebulae.
This nebula does not show the hollow shell structure, but has a greater
equatorial enhancement along the minor axis (P.A.\ of $149^{\circ}$).
These peculiarities have been attributed to a higher degree of
equatorial enhancement in the shell and to the inclination angle 
of the nebula that is marginally edge-on.
However, our $\theta_{\rm eq}$ measurement ($101^{\circ}$), which is not 
consistent with this interpretation, appears affected by the bright
starlight in this marginally optically thin nebula.  

\subsection{Discussion}

\subsubsection{Imaging-Polarimetric Characteristics of DUPLEX PPNs}

Having reviewed the imaging-polarimetric characteristics of individual
sources in \S 3.2, it now appears that the most prevalent  
characteristic among the observed DUPLEX PPNs is their point-symmetric
appearance in the polarized surface brightness.
This point-symmetric appearance of these nebulae in polarized light
stems from the fact that one side of a lobe shows stronger polarization
than the other and that the stronger polarization occurs at the opposite 
sides of the opposite lobes.

Among 13 DUPLEX PPNs in this compilation, 10 sources exhibit a sign
of point-symmetry while the degree of point-symmetry varies among the
sources. 
The exceptions are IRAS 09452, 10131, and 12419.
In IRAS 10131 and 12419, the structure of the polarized surface
brightness is not determined well since the polarized nebula is too
compact in the former and the polarization flux is too weak in the latter.
Thus, neglecting the two inconclusive cases, we have 10 out of 11
objects that display point-symmetric imaging-polarimetric
characteristics.
In IRAS 09452, the main lobe pair (the NE and S lobes in Fig.\
\ref{09452}) is aligned along a straight line and so do the other lobes
(the N and NW lobes) and there is no apparent indication of
point-symmetric variation of polarization in the lobes. 
Thus, IRAS 09452 provides an example in which the imaging-polarimetric
characteristics of the nebula are determined by the illumination effects.
In fact, the N and NW lobes have been shown recently to be due to
scattering of stray starlight leaking through fissures in the optically 
thick toroidal core of the CSE \citep{murakawa05}.
None of these 10 DUPLEX PPNs that possess point-symmetry, however, shows
obvious evidence for the illumination effects as does IRAS 09452. 

In DUPLEX PPNs, the equatorial regions can be very optically thick in
general as evidenced by the presence of a dust lane. 
Nevertheless, the bipolar lobes themselves are optically thin enough
that they are seen via dust-scattered starlight.  
This allows us to adopt the same argument for SOLE PPNs that we used in
Paper I and to investigate the structure of the lobes in DUPLEX PPNs by
relating the {\ipol} and $P$ surface brightnesses to the amount 
of dust along the line of sight.
Following a simple scattering angle effect (i.e., high degrees of
polarization are expected from scattering at near 90$^{\circ}$ angles) 
as in Paper I, we can attribute the observed point-symmetric appearance 
in these 10 DUPLEX PPNs to the column density variations along the
periphery of these nebulae.
Since there is more dust on one side of a lobe than the other, 
we conclude that there is an azimuthal density gradient in these
lobes.       
Asymmetric illumination can potentially induce similar point-symmetric
nebula morphologies in polarized flux.  
However, polarized flux enhancements can occur anywhere and do not have
to occur at the periphery of the nebulae.
This is inconsistent with the observed trend in that the regions of 
enhanced polarization occurs near the periphery of the nebulae.
Thus, we conclude that the asymmetric effect is less likely than the
density effect as the cause for the point-symmetric appearance of the
nebulae in polarized flux.

While the overall imaging-polarimetric morphology of the majority of the
sample sources can be described as point-symmetric, there appear to be
three distinct CSE structures among those point-symmetric nebulae. 
There are nebulae that show polarized fluxes near the periphery of the
lobes in both {\ipol} and $P$ (IRAS 10178, 16594, 17150, 17423, and
19343). 
As discussed for SOLE PPNs in Paper I, this structure suggests that the
lobes are hollow shells.
Such hollow lobes can be created by either the cessation of mass loss in
the polar directions or the sweeping up of the mass-loss ejecta into the
polar directions.
Recently, \citet{arrendondo04} have shown hydro-dynamically that
interactions between an AGB wind and a jet from an accretion disk can
create a narrow-waist bipolar nebula if the jet dominates.
A famous well-collimated bipolar PPN, M2-9 (e.g.\ \citealt{schwarz97}),
which is very similar to IRAS 10178, has been interpreted as a symbiotic
system evolving into a PN based on a scenario in which outflows emanate
from the accretion disk \citep{livio01}.      
In fact, IRAS 19343 has been recently suggested to be a symbiotic system
based on a detailed spectroscopic study \citep{arrieta05}. 
Furthermore, the presence of precessing jets has been suggested in IRAS
17423 and 19343 based on the presence of shock excited knots
\citep{bujarrabal98b,ss01,riera03}. 
Thus, such systems may reasonably explain the morphologies of these
objects and even the oblique angle of the equatorial plane suspected in
some of these objects (see below) may well be explained under the
presence of precessing jets.

There are also nebulae that show $P$ only near the periphery of the
lobes but {\ipol} inside the lobes (IRAS 10197, 17245, and 17441).
These nebulae are also characterized by the very pronounced figure 
{\sf S} shape. 
Unlike the previous case, the hollow $P$ structure does not indicate the
hollowness of the lobes since {\ipol} is detected all over the lobes.
The reduced $P$ in the middle of the lobes is due to increased
unpolarized light in the middle of the lobes.  
Since the central star is not very visible in these sources, it is
not likely that the central star contributes to the increased 
presence of unpolarized light.
Thus, we interpret that the unpolarized component of light is increased
in the middle of these nebulae due to multiple scattering in their 
optically thicker equatorial regions.
Such enhanced optical depths in the vicinity of the central star along
the equatorial plane can be attributed, for example, to the presence of  
a disk around the central star.
It has been shown recently that a warped equatorial density enhancement
can generate point-symmetric lobes by a two-fluid hydro-dynamical
simulations \citep{icke03,rim04}. 
Similarities between models and the nebula structure (especially of IRAS 
10197) are remarkable.  

There are yet another nebulae that display clumpiness in the polarized
surface brightness distribution (IRAS 04395 and AFGL 2688). 
As previously noted, the clumpy structure of IRAS 04395 may be due to
episodic mass loss or instabilities in the outflows.
The bipolar lobes of AFGL 2688 also show signs of hollowness but their
{\ipol} structure (especially of the northern lobe) is clumpy, and
therefore, AFGL 2688 may belong to the previous group. 
However, none of the sources in the previous group shows the presence of
the concentric arcs as in AFGL 2688.
Thus, AFGL 2688 may be a very unique object.
Although the nebula structure is inconclusive due to its compactness,
IRAS 10131 may be related to AFGL 2688 because IRAS 10131 also shows
some arcs in the CSE \citep{schmidt02}.

\subsubsection{SOLE and DUPLEX Morphologies}

In this section we compare the morphological characteristics of SOLE and
DUPLEX PPNs in this the most complete compendium of imaging-polarimetric
data for PPNs taken with {\hst} including data presented in Paper I.
In Paper I, we have shown that (1) optically thin SOLE PPNs consist of a
hollow spheroidal shell with a built-in equatorial density enhancement,
which manifests itself as the toroidal structure and (2) there is a
range of optical depths, which can influence the observed morphological
traits even among SOLE CSEs. 
In the present work, we have concluded that there exist the azimuthal
density gradient in addition to equatorial density enhancements in the bipolar
lobes of optically thicker DUPLEX PPNs.

The distinction between SOLE and DUPLEX PPNs was initially proposed
based primarily on the presence/absence of the optically thick dust lane  
\citep{ueta00}.  
The present imaging polarimetry provides an alternative means to
differentiate these two morphological types.
As we have reviewed in \$ 3.3.1, we have found a prevalent
morphological feature - point-symmetry in the surface brightness
distribution of polarized flux.
This point-symmetric appearance is probably due to the azimuthal density
gradient being reversed in the opposing lobes above and below the
equatorial plane.  
On the other hand, in SOLE PPNs there is no reversal of the azimuthal
density gradient above and below the equatorial plane, although the
azimuthal density gradient 
is present.
The best example for this is IRAS 07134 presented in Paper I, in which  
the presence of the azimuthal density gradient has been shown in the
dust and CO gas distribution in its shell \citep{meixner04}. 

Thus, it appears that mechanisms that generate a reversal in the
azimuthal density gradient in the opposite lobes are related to the
fundamental nature of the DUPLEX nebulae while these mechanisms do not
operate in the SOLE nebulae.
Therefore, we postulate that the emergence of the reversal in the
azimuthal dust density distribution is directly linked to the origin of
the morphological bifurcation between SOLE and DUPLEX PPNs.
However, it is still inconclusive at this point what causes this reversal.

\newpage
\subsubsection{Oblique Orientation of the Equatorial Plane}

In the present work, we have adopted $\theta_{\rm eq}$, the integrated
$\theta$ over an aperture, as a measure of the orientation of the
equatorial plane (modulo $90^{\circ}$ to be precise).
In conventional imaging, the orientation of the equatorial plane is
inferred from the appearance of the dust lane and/or of the bipolar
lobes.
As we have noted earlier, the way starlight gets scattered in the
equatorial regions can be strongly influenced by the local density
enhancements.
Therefore, the silhouette morphology of dust lanes, especially in high
spatial resolution imaging data, can be susceptible to the local density
distribution.
In such cases, imaging polarimetry can alternatively constrain the
orientation of the equatorial plane better than the conventional imaging.

We have seen in \S 3.3.2 that $\theta_{\rm eq}$ does not
necessarily corresponds to the orientation of the apparent dust lane
seen in the $I$ data. 
Among 13 DUPLEX PPNs, we have found in 10 sources that the discrepancy
between $\theta_{\rm eq}$ and the orientation of the bipolar axis is
more than the quoted error, assuming that the uncertainty in the bipolar
axis orientation is $\sim 2^{\circ}$. 
The two angles are equal within errors in one case (IRAS 17150).
In two cases $\theta_{\rm eq}$ is not reliably fixed probably due to the
bright central star and faint polarized surface brightnesses (IRAS 12419
and 16594).  
In the 10 oblique cases, however, the determination of the orientation
of the bipolar axis from the {\ipol} structure is somewhat subjective:
the $2^{\circ}$ uncertainty may be inappropriate for three cases (IRAS
17245, IRAS 17423, and AFGL 2688).   
Even so, there are still seven sources in which the equatorial plane is
more than $\sim10^{\circ}$ oblique from the perpendicular orientation
with respect to the bipolar axis, and there are more than twice as many
oblique cases as orthogonal cases. 

Even if we take a stricter stance on the subjectivity in defining
the bipolar axis and on our informal error analysis in deriving
$\theta_{\rm eq}$, there are cases in which the oblique angle is more
than 20$^{\circ}$ (IRAS 17441 and 19343).
This issue needs to be investigated further in future as it may be
related to the mechanisms to generate the equatorial density enhancement 
and bipolar lobes. 

\section{Summary}

We have presented a compendium of near-IR imaging-polarimetric data 
of evolved star CSEs that have been obtained with {\hst}/NICMOS to date,
in order to study polarization characteristics of these CSEs, especially
of optically thicker bipolar DUPLEX PPNs.
The data have shown that their point-symmetric appearance is prevalent
(seen in 10 out of 11 objects).    
Since the prevalent point-symmetry does not seem to be due to the 
illumination effects in almost all cases (except for one), we have
concluded that the observed point-symmetry is due to the presence of the
azimuthal density gradient in the CSEs. 

These point-symmetric nebulae can be grouped into three based on the
detailed morphologies.
The first group consists of objects that show polarized fluxes near the
periphery of the lobes in both {\ipol} and $P$ (IRAS 10178, 16594,
17150, 17423, and 19343).  
By using the same argument as in Paper I, we suggest that the bipolar
lobes in these sources are hollow shells.
The second group consists of objects that are characterized by the very
pronounced figure {\sf S} shape and show $P$ only near the periphery of
the lobes but {\ipol} inside the lobes (IRAS 10197, 17245, and
17441).  
Since the reduced $P$ in the middle of the nebula is due to the
increased unpolarized light in the nebula, we interpret that 
multiple scattering takes place in the relatively dense equatorial
regions in these sources.
The last group is characterized by their clumpy nebulae (IRAS 04395 and 
AFGL 2688) that may be due to episodic mass loss or instabilities in the
outflows. 
While the lobes of AFGL 2688 show similarities to the second group, the
presence of the concentric arcs makes it rather unique.
IRAS 10131 may be related to AFGL 2688 because of the presence of arcs.

We have found that the ways in which the azimuthal density gradient is
manifested in SOLE and DUPLEX PPNs are very distinct: the direction of
the azimuthal density gradient in DUPLEX PPNs is reversed in the
opposing lobes above and below the equatorial plane whereas such a
reversal is not seen in SOLE PPNs.
The most probable explanation for this seems to be the presence of
precessing outflows into the opposite directions and/or magnetic field
that is reversed its polarity above and below the equatorial plane.
We have also found indications that the equatorial plane of the system
(defined by $\theta_{\rm eq}$) is not necessarily orthogonal to the axis
of the apparent bipolar structure in the total intensity data (at least
7 out of 10 sources). 
There are cases in which the oblique angle is more than 20$^{\circ}$
(IRAS 17441 and 19343). 
Therefore, we have concluded that the occurrence of the oblique
orientation of the equatorial plane with respect to the bipolar axis is
real and that this issue needs to be investigated further in future.

\acknowledgements
This research is based on observations with the NASA/ESA Hubble Space
Telescope, obtained at the Space Telescope Science Institute (STScI),
which is operated by the Association of Universities for Research in
Astronomy, Inc.\ (AURA) under NASA contract No.\ NAS 5-26555.
Authors acknowledge financial support by NASA STI 9377.05-A.
Ueta also acknowledge partial support from the project IAP P5/36
financed by Federal Services for Scientific, Technical and Cultural
Affairs of the Belgian State and the Universities Space Research
Association SOFIA (Stratospheric Observatory for Infrared Astronomy)
office at NASA Ames Research Center as well as a US 
National Research Council Research Associateship Award and a NASA
Postdoctoral Program Award.

\clearpage
\begin{landscape}
\begin{deluxetable}{llccccccrrcc}
\tablecolumns{12} 
\tablewidth{0pt} 
\tabletypesize{\scriptsize}
\tablecaption{\label{data}%
{\hst}/NICMOS Imaging-Polarimetric Observations of Evolved Stars} 
\tablehead{%
\colhead{} &
\colhead{} &
\colhead{} &
\colhead{} &
\colhead{} &
\colhead{} & 
\multicolumn{2}{c}{DITHER} &
\colhead{EXPTIME\tablenotemark{b}} &
\colhead{ORIENTAT\tablenotemark{c}} &
\colhead{} & 
\colhead{}\\
\cline{7-8}
\colhead{Source} &
\colhead{Alternative Names} &
\colhead{Date} &
\colhead{FILTER\tablenotemark{a}} &
\colhead{SAMP\_SEQ} &
\colhead{NSAMP} &
\colhead{PATTERN} &
\colhead{NPTS} &
\colhead{(sec)} &
\colhead{(deg)} &
\colhead{PID} &
\colhead{Reference}} 

\startdata 
\cutinhead{Source Data}
IRAS 02168$-$0312 &
Mira, O Cet &
1998 Aug 11 &
POL-S &
STEP8 &
13 &
NONE &
\dots &
135.8608\phn &
34.7588 &
7416 &
\\
&
&
1998 Aug 11 &
POL-L &
STEP32 &
4 &
TWO-CHOP &
4 &
169.8279\phn &
33.9539 &
7416 &
\\    	
IRAS (z)02229$+$6208\tablenotemark{d} & 
&
2003 Mar 28 & 
POL-S &
STEP8 & 
16 &
SPIRAL &
4 &
319.6832\phn & 
169.035\phn &
9377 &
1 \\
IRAS 04296$+$3429 & 
&
2003 Mar 28 & 
POL-S &
STEP8 & 
19 &
SPIRAL &
5 &
519.5079\phn & 
$-$149.313\phn & 
9377 &
1, 2 \\
IRAS 04395$+$3601 & 
AFGL/CRL 618&
2003 Nov 26 & 
POL-L &
SPARS64 & 	
13 &
SPIRAL &
4 &
2559.868\phn\phn & 
$-$35.4301 & 
9430 &
 \\
IRAS 06530$-$0213 & 
&
2003 Mar 29 & 
POL-S &
STEP8 & 
19 &
SPIRAL &
5 &
519.5079\phn & 
$-$131.793\phn &
9377 &
1, 2 \\
IRAS 07134$+$1005 & 
HD 56126 &
2003 Mar 29 & 
POL-S &
STEP8 & 
19 &
SPIRAL &
5 &
519.5079\phn & 
$-$129.432\phn &
9377 &
1 \\
IRAS 09452$+$1330 &  	
IRC +10 216, CW Leo&
1997 Apr 30 &
POL-S &
STEP8 &
9 &
NONE &
\dots &
71.89689 &
$-$113.462\phn &	
7120 &
3, 4 \\
IRAS 10131$+$3049 &
CIT 6, RW LMi&
1998 May \phn5 &
POL-S &
STEP16 &
12 &
SPIRAL &
4 &
319.7775\phn &
$-$109.901\phn &
7854 &
5 \\
&
&
1998 Oct 11 &
POL-S &
STEP16 &
11 &
SPIRAL &
4 &
255.8052\phn &
80.8889 &
7854 &
5 \\
&
 &
1998 Feb 23 &
POL-L &
MCAMRR &
12 &
SPIRAL &
4 &
3.32561 &
$-$52.5125 &
7854 &
5 \\
&
 &
1998 May \phn5 &
POL-L &
MCAMRR &
12 &
SPIRAL &
4 &
3.32561 &
$-$110.707\phn &
7854 &
5 \\
&
 &
1998 Oct 11 &
POL-L &
MCAMRR &
11 &
SPIRAL &
4 &
3.02328 &
80.0841 &
7854 &
5 \\
IRAS 10178$-$5958 &
&
1998 Mar \phn6 &
POL-L &
STEP32 &
17 &
SPIRAL &
3  &
863.8767\phn &
124.507\phn &
7840 &
\\
IRAS 10197$-$5750 &
Hen 3-404, Roberts 22&
1998 May 18 &
POL-L &
STEP32 &
12 &
SPIRAL &
3 &
383.8884\phn &
$-$133.434\phn &
7840 &
\\
IRAS 12419$-$5414 &
Boomerang Nebula&
1998 Sep 12 &
POL-L &
STEP32 &
17 &
SPIRAL &
3 &
863.8767\phn &
$-$65.4931 &
7840 &
\\
IRAS 16594$-$4656 &
&
1998 May \phn2 &
POL-L  &
STEP32 &
17 &
SPIRAL &
3 &
863.8767\phn &
78.9531 &
7840 &
1, 6 \\
IRAS 17150$-$3224 &
&
1998 Aug 16 &
POL-L &
STEP64 &
16 &
SPIRAL &
3 &
1343.842\phn\phn &
$-$150.493\phn &
7840 &
6 \\
IRAS 17245$-$3951 &
OH 348.81$-$2.84&
1998 Apr 30 &
POL-L  &
STEP16 &
24 &
SPIRAL &
3 & 
815.5839\phn &
64.7606 &
7840 &
6, 7 \\
IRAS 17423$-$1755 &
Hen 3-1475&
1997 Aug \phn9 &
POL-S &
STEP8 &
21 &
SPIRAL &
2 &
239.7774\phn &
$-$137.381\phn &
7285 &
\\
IRAS 17441$-$2411 &
&
1998 Mar 10 &
POL-L &
STEP32 &
21 &
SPIRAL &
3 &
1247.867\phn\phn &
45.665\phn &
7840 &
6\\
IRAS 19343$+$2926 &
M 1-92&
1998 Apr 24 &
POL-L &
STEP128 &
12 &
XSTRIP &
3 &
767.8638\phn &
29.2312 &
7839 &
\\
AFGL 2688 &
Egg Nebula, V1610 Cyg&
1998 Apr \phn8 &
POL-S &
STEP16 &
20 &
SPIRAL &
4 &
831.556\phn\phn &
58.3121 &
7423 &
3 \\
&
&
2002 Oct \phn6 &
POL-S &
STEP16 &
20 &
SPIRAL &
4 &
831.556\phn\phn &
$-$117.002\phn &
9644 &
\\
&
&
1997 Apr \phn8 &
POL-L &
STEP16 &
26 &
SPIRAL &
4 &
1215.39\phn\phn\phn &
56.3451 &
7115 &
8, 9 \\
&
&    	
2002 Oct \phn6 &
POL-L & 
STEP16 &
26 &
SPIRAL &
4 &
1215.39\phn\phn\phn &
$-$117.752\phn &
9644 &
\\
\cutinhead{PSF Calibration Source Data}
HD 12088 & 
&
2003 Mar 28 & 
POL-S &
STEP1 & 
19 &
SPIRAL &
5 &
79.7868\phn &
162.51\phn\phn & 
9377 &
1 \\
BD $+$32$^{\circ}$3739 & 
&
1997 Sep \phn1 & 
POL-S &
SCAMRR & 
9 &
NONE &
\dots &
11.368\phn\phn  &
$-$92.7652 &
7692 &
1, 3 \\
&
 & 
2002 Sep \phn9 & 
POL-S &
SCAMRR & 
9 &
SPIRAL &
4 &
12.992\phn\phn  &
$-$108.88\phn\phn &
9644 & 
1\\
&
 &
1997 Sep \phn1 & 
POL-L &
STEP1 & 
9 &
NONE &
\dots &
41.86784  &
$-$93.5733 &
7692 &
1, 3 \\
&
 & 
2003 Jun \phn8 & 
POL-L &
STEP1 & 
9 &
SPIRAL &
4 &
47.84896  &
$-$109.63\phn\phn &
9644 &
1 \\
\enddata
\tablenotetext{a}{POL-S: short wavelength polarizers ($0.8 - 1.3\micron$, centered at $1.1\micron$); 
POL-L: long wavelength polarizers ($1.89 - 2.1\micron$, centered at 2.05\micron).}
\tablenotetext{b}{Total exposure time per polarizer.}
\tablenotetext{c}{The ORIENTAT header parameter refers to PA of the image +y axis (degrees E of N).}
\tablenotetext{d}{The ``z'' prefix in the IRAS designation is given 
to indicate the fact that this object was found in the Faint Source 
Reject File in the {\iras} Faint Source Survey \citep{hrivnak99}.}
\tablerefs{%
1.\ \citet{ueta05},
2.\ \citet{gledhill05},
3.\ \citet{hines00},
4.\ \citet{oster00},
5.\ \citet{schmidt02},
6.\ \citet{su03},
7.\ \citet{gledhill01},
8.\ \citet{sahai98},
9.\ \citet{weintraub00}}
\end{deluxetable} 
\clearpage
\end{landscape}

\clearpage
\begin{landscape}
\begin{deluxetable}{lcccccrrccccc}
\tablecolumns{13} 
\tablecaption{\label{results}%
Summary of {\hst}/NICMOS Imaging Polarimetry of the Circumstellar Shell
 around Evolved Stars} 
\tablehead{%
\colhead{} & 
\multicolumn{3}{c}{Measured Coordinates (J2000)\tablenotemark{a}} &
\colhead{} &
\multicolumn{3}{c}{Flux} &
\colhead{} &
\colhead{} &
\multicolumn{3}{c}{Polarized Surface Brightness Distribution} \\
\cline{2-4}
\cline{6-8}
\cline{11-13}
\colhead{} &
\colhead{} &
\colhead{} &
\colhead{Offset} &
\colhead{} &
\colhead{} &
\colhead{$I$} &
\colhead{{\ipol}} &
\colhead{$P_{\rm max}$\tablenotemark{c}} &
\colhead{$\left<P\right>$\tablenotemark{c}} &
\colhead{} &
\colhead{Size} &
\colhead{P.A.\tablenotemark{e}} \\
\colhead{Source} &
\colhead{R.A.} &
\colhead{Decl.} &
\colhead{(arcsec)} &
\colhead{} &
\colhead{Band\tablenotemark{b}} &
\colhead{(mJy)} &
\colhead{(mJy)} &
\colhead{($\%$)} &
\colhead{($\%$)} &
\colhead{Morphology\tablenotemark{d}} &
\colhead{(arcsec)} &
\colhead{(deg)}} 

\startdata
IRAS 02229 & 
02 26 41.79 & 
$+$62 21 22.2 & 
$0.12\pm0.31$ & 
 & 
POL-S &
4800 & 
1200 &
60 &
$34\pm11$ &
ELS &
$2.1 \times 1.3$ & 
\phn59 \\

IRAS 04296 & 
04 32 56.95 & 
$+$34 36 13.1 & 
$0.06\pm0.17$ & 
 & 
POL-S &
340 & 
\phn\phn60 &
84 &
$29\pm13$ &
Dual ELS (hollow) &
$2.1 \times 0.7$ & 
\phn26 \\
&
&
&
&
&
&
&
&
&
&
&
$3.5 \times 0.5$ &
\phn99 \\

IRAS 04395 &
04 42 53.61 &
$+$36 06 53.8 &
$0.23\pm0.16$ &
 & 
POL-L &
275 &
38 &
54 &
$19\pm10$ &
BPL (inverse {\sf S}) &
$15.0 \times 3.4$ &
\phn94 \\

IRAS 06530 & 
06 55 31.80 & 
$-$02 17 28.3 & 
$0.04\pm0.20$ & 
 & 
POL-S &
\phn380 & 
\phn\phn67 &
68 &
$33\pm14$ &
ELS $+$ Cusp (hollow) &
$2.7 \times 1.0$ &
\phn20 \\

IRAS 07134 & 
07 16 10.27 & 
$+$09 59 48.5 & 
$0.08\pm0.27$ &
 & 
POL-S &
9300\tablenotemark{f} & 
2000 &
81 &
$44\pm20$ &
ELS (hollow) &
$4.8 \times 4.0$ &
\phn25 \\

IRAS 09452 &
09 47 57.37 &
$+$13 16 42.8 &
$0.04\pm0.13$ &
 & 
POL-S &
4590 & 
808  &
65 &
$28\pm12$ &
Fragmented BPL &
$5.4 \times 2.2$ &
\phn22 \\

IRAS 10131 (1998-05-05) &
10 16 02.26 &
$+$30 34 18.8 &
$0.06\pm0.09$ &
 & 
POL-S &
2360 &
465 &
60 &
$26\pm14$ &
BPS &
$10.0 \times 3.6$ &
\phn35 \\
\phm{IRAS 10131} (1998-10-11) &
10 16 02.19 &
$+$30 34 18.6 &
$0.08\pm0.16$ &
 & 
POL-S &
8470 &
1540 &
60 &
$25\pm13$&
 &
 &
\phn35 \\

IRAS 10178 &
10 19 32.48 &
$-$60 13 28.8 &
$0.09\pm0.07$ &
 & 
POL-L &
554 &
146 &
56 &
$27\pm12$ &
BPL ({\sf S})&
$16.4 \times 2.6$ &
\phn73 \\

IRAS 10197 &
10 21 33.88 &
$-$58 05 47.7 &
$0.26\pm0.11$ &
 & 
POL-L &
1320 &
330 &
65 &
$29\pm14$ &
BPS ({\sf S}) $+$ Protrusions &
$7.6 \times 5.6$ &
\phn23 \\

IRAS 12419 &
12 44 46.07 &
$-$54 31 13.5 &
$0.06\pm0.25$ &
 & 
POL-L &
380 &
35 &
41 &
$12\pm\phn8$ &
BPS ({\sf X}) &
$10.0 \times 7.0$ &
163 \\

IRAS 16594 & 
17 03 10.04 & 
$-$47 00 27.0 &
$0.06\pm0.14$ & 
 & 
POL-L &
\phn620 & 
\phn\phn66 &
73 &
$39\pm16$ &
ELS $+$ Cusp &
$5.0 \times 2.2$  &
\phn82 \\

IRAS 17150 &
17 18 19.81 &
$-$32 27 21.2 &
$0.14\pm0.07$ &
 & 
POL-L &
\phn230 &
\phn\phn60 &
70 &
$26\pm18$ &
BPL ({\sf S}) &
$6.6 \times 2.2$ &
$122$ \\

IRAS 17245 &
17 28 04.58 &
$-$39 53 45.1 &
$0.24\pm0.08$ &
 & 
POL-L &
\phn145 &
\phn\phn18 &
48 &
$17\pm11$ &
BPL &
$2.0 \times 1.3$ &
\phn15 \\

IRAS 17423 &
17 45 14.23 &
$-$17 56 47.0 &
$0.15\pm0.11$ &
 & 
POL-S &
\phn272\tablenotemark{g} &
\phn\phn48\tablenotemark{g} &
51 &
$26\pm12$ &
BPL ({\sf S}) $+$ Knots & 
$10.9\tablenotemark{g} \times 1.7$ &
133 \\

IRAS 17441 &
17 47 13.51  &
$-$24 12 52.0 &
$0.19\pm0.16$ &
 & 
POL-L &
\phn212 &
\phn\phn57 &
59 &
$16\pm14$ &
BPL ({\sf S}) &
$4.0 \times 2.1$ &
\phn\phn1 \\

IRAS 19343 &
19 36 19.02 &
$+$29 32 50.2 &
$0.27\pm0.22$ &
 & 
POL-L &
3360 &
373 &
52 &
$22\pm13$ &
BPL ({\sf S}) &
$12.9 \times 3.9$ &
131 \\

AFGL 2688 (1998-04-08) &
21 02 18.66 &
+36 41 37.4 &
$\phm{0.00\pm}0.17$\tablenotemark{h} &
 & 
POL-S &
\phn413\tablenotemark{g} &
\phn226\tablenotemark{g} &
89 & 
$50\pm18$ &
BPL (inverse {\sf S}) &
$15.0\tablenotemark{g} \times 4.0$ &
\phn12 \\
\phm{AFGL 2688} (2002-10-06) &
21 02 18.53 &
+36 41 36.2 &
$\phm{0.00\pm}0.23$\tablenotemark{h} &
 & 
POL-S &
469\tablenotemark{g} &
272\tablenotemark{g} &
81 &
$51\pm15$ &
 \\
\phm{AFGL 2688} (1997-04-08) &
21 02 18.76 &
$+$36 41 36.9 &
$\phm{0.00\pm}0.29$\tablenotemark{h} &
 & 
POL-L &
\phn599\tablenotemark{g} &
\phn284\tablenotemark{g} &
74 &
$23\pm17$ &
BPL (inverse {\sf S}) &
$23.7\tablenotemark{g} \times 3.7$ &
\phn12 \\
\phm{AFGL 2688} (2002-10-06) &
21 02 18.55 &
$+$36 41 36.0 &
$\phm{0.00\pm}0.22$\tablenotemark{h} &
 & 
POL-L &
\phn637\tablenotemark{g} &
\phn295\tablenotemark{g} &
73 &
$23\pm16$ &
 \\
\enddata
\tablenotetext{a}{Of the centrosymmetric polarization pattern center.
Offset is the angular distance from the pattern center to the $I$ peak
 with an associated error in the pattern center determination.}
\tablenotetext{b}{POL-S: short wavelength polarizers ($0.8 -
 1.3\micron$, centered at $1.1\micron$); POL-L: long wavelength
 polarizers ($1.89 - 2.1\micron$, centered at 2.05\micron).} 
\tablenotetext{c}{$P_{\rm max}$ is the maximum polarization strength
 detected, while $\left<P\right>$ refers the mean $P$ and its standard
 deviation.} 
\tablenotetext{d}{ELS: elliptical shell; BPS: bipolar shell without
 any apparent dust lane; BPL: bipolar lobes separated by a dust lane.}
\tablenotetext{e}{Of the nebula (counter-clockwise east of north).}
\tablenotetext{f}{The measured flux value is three times higher 
than the previously observed value (see \S 4 in Paper I).}
\tablenotetext{g}{The observed field did not cover the entire nebula.}
\tablenotetext{h}{No $I$ peak.}
\end{deluxetable} 
\clearpage
\end{landscape}

\begin{deluxetable}{lccrcc}
\tablecolumns{6} 
\setlength{\tabcolsep}{0.03in}
\tabletypesize{\tiny}
\tablecaption{\label{aperture}%
Summary of Aperture Polarimetry}
\tablehead{%
\colhead{} & 
\colhead{} & 
\colhead{Aperture} &
\colhead{$P$\tablenotemark{a}} &
\colhead{$\theta_{\rm eq}$\tablenotemark{a}} &
\colhead{P.A.\tablenotemark{b}} \\
\colhead{Source} &
\colhead{Band} &
\colhead{(arcsec)} &
\colhead{($\%$)} &
\colhead{(deg)} &
\colhead{(deg)}}

\startdata
IRAS 02229 & 
POL-S &
0.4 &
$6.1\pm1.0$ & 
$101\pm1$ &
59 \\

IRAS 04296 & 
POL-S &
0.6 &
$5.7\pm1.0$ & 
$11\pm1$ &
26, 99 \\

IRAS 04395 &
POL-L &
1.2 &
$7.7\pm0.5$ &
$83\pm1$ &
94 \\

IRAS 06530 & 
POL-S &
1.6 &
$2.7\pm1.5$ & 
$159\pm2$ &
20 \\

IRAS 07134 & 
POL-S &
0.6 &
$4.9\pm2.0$ & 
$136\pm1$ &
25 \\

IRAS 09452 &
POL-S &
0.8 &
$11.2\pm1.0$ & 
$32\pm1$ &
22 \\

IRAS 10131 (1998-05-05) &
POL-S &
0.6 &
$4.6\pm0.7$ & 
$49\pm1$ &
35 \\
\phm{IRAS 10131} (1998-10-11) &
POL-S &
0.6 &
$5.4\pm0.7$ & 
$43\pm1$ &
35 \\

IRAS 10178 &
POL-L &
1.0 &
$3.0\pm1.5$ & 
$80\pm1$ &
73 \\

IRAS 10197 &
POL-L &
3.0 &
$8.5\pm1.0$ & 
$34$\tablenotemark{c} &
23 \\

IRAS 12419 &
POL-L &
0.9 &
$3.5\pm1.0$ & 
$57$\tablenotemark{c} &
163 \\

IRAS 16594 & 
POL-L &
0.9 &
$0.7\pm0.2$ & 
$22\pm6$ &
82 \\

IRAS 17150 &
POL-L &
2.4 &
$17.5\pm2.0$ & 
$120\pm1$ &
122 \\

IRAS 17245 &
POL-L &
1.1 &
$7.1\pm0.3$ & 
$21$\tablenotemark{c} &
15 \\

IRAS 17423 &
POL-S &
2.2 & 
$5.2\pm0.1$ & 
$128\pm2$ &
133 \\

IRAS 17441 &
POL-L &
1.0 &
$11.7\pm1.0$ & 
$28\pm4$ &
1 \\

IRAS 19343 &
POL-L &
0.9 &
$2.3\pm4.0$ & 
$151$\tablenotemark{c} &
131 \\

AFGL 2688 (1997-04-08) &
POL-L &
3.8 &
$42.5\pm1.0$ &
$19\pm1$ &
12 \\
\phm{AFGL 2688} (2002-10-06) &
POL-L &
3.0 &
$37.3\pm0.8$ & 
$17\pm1$ &
12 

\enddata
\tablenotetext{a}{These errors are not via a formal error analysis. See
 text for details.}  
\tablenotetext{b}{Of the nebula (counter-clockwise east of north).
 Typically $\pm2^{\circ}$ error. Reproduced from Table \ref{results}.}
\tablenotetext{b}{Error less than $0.5\deg$.}
\end{deluxetable} 

\clearpage

\clearpage

\clearpage

\clearpage

\clearpage

\clearpage

\clearpage

\clearpage

\clearpage

\clearpage

\clearpage

\clearpage

\clearpage

\clearpage

\clearpage

\end{document}